\def\OX{\Omega_\textrm{X}}
\def\IX{I_\textrm{X}}
\def\GMX{\mu_\textrm{X}}
\def\aX{a_\textrm{X}}
\def\eX{e_\textrm{X}}
\def\masy{~\textrm{mas~yr}^{-1}}
\def\deg{~\textrm{deg}}
\def\eqt{_\textrm{eq}}

\def\kx{{\hat{S}}_x}
\def\ky{{\hat{S}}_y}
\def\kz{{\hat{S}}_z}

\def\nk{n_{\rm b}}

\def\Pb{P_{\rm b}}

\def\rfr#1{Equation~(\ref{#1})}
\def\rfrs#1#2{Equations~(\ref{#1})~to~(\ref{#2})}

\def\virg#1{``#1"}

\def\eqi{\begin{equation}}
\def\eqf{\end{equation}}
\def\eqia{\begin{eqnarray}}
\def\eqfa{\end{eqnarray}}

\def\rp#1#2{{#1\over#2}}
\def\lb#1{\label{#1}}

\def\bds#1{\boldsymbol{#1}}

\def\ton#1{\left(#1\right)}
\def\qua#1{\left[#1\right]}
\def\grf#1{\left\{#1\right\}}
\def\ang#1{\left\langle #1\right\rangle}
\documentclass[onecolumn]{aastex}

\usepackage[title]{appendix}
\usepackage{hyperref}
\usepackage{booktabs}
\usepackage[table,xcdraw]{xcolor}
\usepackage{multirow}
\usepackage{rotating,tabularx}
\usepackage{float}
\usepackage{enumerate}
\usepackage[polutonikogreek,english]{babel}
\usepackage{amsmath,textgreek,w-greek,wasysym}
\usepackage{amsthm}
\usepackage{amscd,lineno}
\usepackage{amssymb,dsfont}
\usepackage{graphicx,epsfig}
\usepackage{txfonts}
\usepackage{xr-hyper}

\RequirePackage{color}

\newcommand{\emaila}{lorenzo.iorio@libero.it}
\newcommand{\grk}[1]{\selectlanguage{polutonikogreek}
#1\selectlanguage{english}}

\allowdisplaybreaks[1]

\begin{document}

\title{Measuring general relativistic dragging effects in the Earth's gravitational field with ELXIS: a proposal}

\shortauthors{L. Iorio}

\author{Lorenzo Iorio\altaffilmark{1} }
\affil{Ministero dell'Istruzione, dell'Universit\`{a} e della Ricerca
(M.I.U.R.)-Istruzione
\\ Permanent address for correspondence: Viale Unit\`{a} di Italia 68, 70125, Bari (BA),
Italy}

\email{\emaila}

\begin{abstract}
In a geocentric kinematically rotating ecliptical coordinate system in geodesic motion through the deformed spacetime of the Sun, both the longitude of the ascending node $\Omega$ and the inclination $I$ of an artificial satellite of the spinning Earth are affected by the post-Newtonian gravitoelectric De Sitter and gravitomagnetic Lense-Thirring effects. By choosing a circular  orbit with $I = \Omega = 90\deg$ for a potential new spacecraft, which we propose to name ELXIS, it would be possible to measure each of the gravitomagnetic precessions separately at a percent level, or, perhaps,  even better depending on the level of accuracy of the current and future global ocean tide models since the competing classical long-term perturbations on $I,~\Omega$ due to the even and odd zonal harmonics $J_\ell,~\ell=2,~3,~4,\ldots$ of the geopotential ideally vanish. Moreover, a suitable linear combination of $I,~\Omega$ would be able to cancel out the solid and ocean tidal perturbations induced by the $K_1$ tide and, at the same time,  enforce the geodetic precessions yielding a secular trend of $-8.3~\textrm{milliarcseconds~per~year}$, thus strengthening the goal of a $\simeq 10^{-5}$ test of the De Sitter effect recently proposed in the literature in the case of an equatorial coordinate system. Relatively mild departures $\Delta I = \Delta\Omega\simeq 0.01-0.1\deg$ from the ideal orbital configuration with $I = \Omega = 90\deg$ are allowed. Present-day levels of relative accuracy in testing the geodetic and the gravitomagnetic effects in the field of the Sun and the Earth, respectively, are $6.4\times 10^{-3}$ (Lunar Laser Ranging) and $3\times 10^{-3}$ (Gravity Probe B) for the De Sitter precessions, and $1.9\times 10^{-1}$ for the Pugh-Schiff rates of change of gyroscopes (Gravity Probe B). Other tests of the Lense-Thirring effect with the LAGEOS type satellites are ongoing in the field of the Earth; their overall accuracy is currently debated.
\end{abstract}

keywords{
Experimental studies of gravity; Experimental tests of gravitational theories; Satellite orbits; Harmonics of the gravity potential field
}

%

\section{Introduction}\lb{intro}
\citet{2018arXiv180901730I} recently proposed to use a hypothetical new terrestrial artificial satellite,  here dubbed\footnote{From \grk{<'elxis}, which means `dragging', `trailing'.} ELXIS and to be placed in a circular path in an orbital plane displaced by $\Omega\eqt = 90\deg$ with respect to the reference direction of the Vernal Equinox $\aries$ perpendicularly to the Earth's equator, in order to measure the general relativistic De Sitter effect \citep{1916MNRAS..77..155D,1918KNAB...27.214S,1921KNAB...23..729F} on the orbital inclination $I\eqt$ to the equator \citep{2018arXiv180901730I} with a possible relative accuracy level of $\simeq 10^{-5}$. A rather strict polar orbital configuration, with departures as little as $\Delta I\eqt\simeq 10^{-3}-10^{-5}\deg$, would be required to reduce the impact of the aliasing perturbations due to the solid and ocean components of the $K_1$ tide, which would be one of the major sources of systematic errors, especially if not too high altitudes were to be adopted. The long-term rates of change of $I\eqt$ due to the even and odd zonal harmonics of the geopotential vanish for the orbital geometry proposed.  It was tacitly assumed that the data analysis would be performed in a geocentric kinematically rotating and dynamically non-rotating \textcolor{black}{\citep{1989NCimB.103...63B,1994PhRvD..49..618D,2003AJ....126.2687S,2011rcms.book.....K}} coordinate system having the mean Earth's equator at the reference epoch J2000.0 as reference $\grf{x,~y}$ plane, and all the angular orbital elements in \citet{2018arXiv180901730I} are to be intended as referred to it.
In the \textcolor{black}{standard} satellite data reductions \textcolor{black}{performed for a variety of purposes}, a kinematically non-rotating and dynamically rotating
geocentric equatorial coordinate system, the International Celestial Reference System (ICRS), is routinely used, i.e. the De Sitter precession is accounted for. \textcolor{black}{It is dynamically rotating because of the fictitious forces in the satellite's equations of motion arising from the rotation required to compensate the kinematic De Sitter precession with respect to distant quasars. They are included in the data processing algorithms in accordance with the IERS Standards \citep{2003AJ....126.2687S,2010ITN....36....1P}; see also \citet[pag.~409]{2011rcms.book.....K}.}

In this paper, we show that, by using a\footnote{Here and in the following, it is assumed that it is also dynamically non-rotating.} kinematically rotating  geocentric coordinate system with the mean ecliptic at J2000.0 as reference $\grf{x,~y}$ plane, which preserves the same satellite's orbital geometry of \citet{2018arXiv180901730I}, it is possible to suitably combine the (ecliptical) node $\Omega$ and  inclination $I$, both affected by the De Sitter precessions \citep{2018arXiv180901730I}, in order to cancel out, by construction, the effect of both the solid and ocean perturbations due to the $K_1$ tide and produce an overall De Sitter secular trend of about $-8.3\masy$. Such a combination would  be impacted neither by the zonal $055.565$ tide nor by the zonals of the geopotential whose perturbations on $I$ and $\Omega$ ideally vanish for $I = \Omega = 90\deg$, thus enforcing the goal of reaching a $\simeq 10^{-5}$ level. Furthermore, it would also be possible to analyze the node and the inclination separately to measure the general relativistic Lense-Thirring effect \citep{1918PhyZ...19..156L} affecting each of them \citep{2011PhRvD..84l4001I} to a few percent accuracy, or, in perspective, even better, depending on the accuracy of the present and future global ocean tide models adopted.
The approach proposed in the present paper would allow to somewhat relax the strict conditions on $I,~\Omega$ also for relatively low orbits with respect to
\citet{2018arXiv180901730I}.

\textcolor{black}{In putting into context the level of accuracy, in principle, obtainable with the presently proposed scenario and assessing its importance properly, the following considerations are in order. The most recent measurement of the geodetic precession was obtained with the Lunar Laser Ranging (LLR) technique \citep{1994Sci...265..482D} for the motion of the Earth-Moon system in the field of the Sun with an accuracy level of $9\times 10^{-4}$  \citep{2018CQGra..35c5015H}. However, it is important to remark that such a figure is likely too optimistic because of an analysis of the systematic errors which might be neither reliable nor robust, as pointed out by \citet{2018CQGra..35c5015H} themselves at the end of their Sect.~4.4. Previously, \citet{2004PhRvL..93z1101W} obtained a relative accuracy of $6.4\times 10^{-3}$ level with LLR. The past Gravity Probe B (GP-B) mission  measured the geodetic precession of four orbiting man-made gyroscopes in the gravitational field of the Earth, reaching an overall relative accuracy of $3\times 10^{-3}$ in a dedicated spaceborne experiment \citep{2011PhRvL.106v1101E,2015CQGra..32v4001E}.}
\textcolor{black}{As far as the gravitomagnetic field of the Earth is concerned, GP-B  measured also the Pugh-Schiff precessions \citep{Pugh59,Schiff60} of the onboard gyroscopes to an accuracy level of $19\%$  \citep{2011PhRvL.106v1101E,2015CQGra..32v4001E}. At present, no aspects of the GP-B results have been criticized in the published literature; on the other hand, the obtained accuracy is not as good as the originally expected one, which was of the order of $1\%$ \citep{2001LNP...562...52E}. Other tests of the gravitomagnetic field of the Earth have been performed in the last twenty years, and are still ongoing, with the geodetic satellites of the LAGEOS type whose Lense-Thirring orbital precessions have been measured with the Satellite Laser Ranging (SLR) technique with increasing accuracy over the years \citep{2012EPJP..127..133C,2013NuPhS.243..180C,2010ASSL..367..371C,2016EPJC...76..120C}. Nonetheless, some aspects of them have been criticized in the literature so far, and their accuracies is the subject of a lingering debate \citep{2009SSRv..148...71C,2012NewA...17..341C,2011Ap&SS.331..351I,2011EL.....9630001I,2017EPJC...77...73I,2012CaJPh..90..883R,2013CEJPh..11..531R,2013NewA...23...63R,2014NewA...29...25R,2015AcAau.113..164R}. For further planned and ongoing SLR-based investigations of the Lense-Thirring effect with the LAGEOS type satellites within the LARASE program, see  \citet{2015CQGra..32o5012L,2016AdSpR..57.1928V,2018PhRvD..98d4034V,2018CeMDA.130...66P}.}

The paper is organized as follows.
In Section~\ref{tras}, a general scheme for obtaining the rates of change of the satellite's inclination and node in the ecliptical coordinate system from the equatorial one is outlined. In Section~\ref{prece}, the long-term  effects on $I,~\Omega$ due to the general relativistic Lense-Thirring effect (Section~\ref{LTrate}) and the odd and even zonal harmonics $J_\ell,~\ell=2,~3,~4,\ldots$ of the Earth's geopotential (Section~\ref{zonrate}) are analytically and numerically worked out. It is shown that departures of $\simeq 0.01-0.1\deg$ from the ideal condition $I = \Omega = 90\deg$ would affect the mismodelled classical precessions to less than the percent level of the gravitomagnetic ones even by assuming very conservative uncertainties in the zonals themselves. Section~\ref{tides} is devoted to the tidal perturbations induced on $I,~\Omega$ by the solid (Section~\ref{soltid}) and ocean (Section~\ref{octid}) components of the $K_1$ tide for $\ell=2,~m=1,~p=1,~q=0$ and their sensitivity to departures of the actual satellite's inclination and nodes from the nominal scenario $I = \Omega = 90\deg$ (Section~\ref{senstid}). It is shown that the largest nominal perturbations arise from the ocean tide;  depending on the accuracy of the latest global ocean tide models, their impact on the Lense-Thirring rates may be as low as a few percent. A linear combination of the precessions of $I,~\Omega$ able to cancel out the $K_1$ tidal perturbations is designed (Section~\ref{combtid}). Unfortunately, it would remove also the Lense-Thirring rates as well. In Section~\ref{des}, it is shown that, instead, the De Sitter precessions are not canceled out by the aforementioned linear combination. The impact of the 3rd-body perturbations due to a distant perturber such as the Moon on both the individual precessions of the inclination and the node and their linear combination is treated in Section~\ref{3rd}. In view of the present-day level of mismodeling in the lunar gravitational parameter, it turns out that the combined De Sitter trend would be affected, at most, at the $\simeq 3\times 10^{-5}-1\times 10^{-4}$ level, while the bias on the Lense-Thirring precessions taken individually would be negligible.  A cursory overview of the impact of the non-gravitational perturbations on both the individual Lense-Thirring precessions and the combined De Sitter effect is given in Section~\ref{ngp}. By relying upon Sec.~(6) of \citet{2018arXiv180901730I} for the inclination and on several works by other researchers for the node, it turns out that, for a geodetic satellite of \textcolor{black}{LAGEOS type}, their effect can be deemed as negligible with respect to the accuracy goal in the proposed relativistic tests. In Section~\ref{vanpa}, we offer a comparison with the past proposal by \citet{1976PhRvL..36..629V} encompassing the launch of two drag-free counter-orbiting spacecraft in nearly identical circular polar orbits. Apart from being simpler and cheaper since it involves the use of a single satellite, ELXIS would be much more accurate, especially as far as the De Sitter effect is concerned. Furthermore, it would bear the possibility of increasing accuracy in forthcoming tests in view, once in orbit, of future improvements in measurement and modeling. Our findings and conclusions are resumed in Section~\ref{fine}.  \textcolor{black}{For the benefit of the reader,} Appendix~\ref{appena} displays a list of definitions of all the physical and orbital parameters used in the text, while  Appendix~\ref{appenb} contains  the numerical values of most of them along with the figures.
\section{The rates of change of the inclination and the node in the ecliptic coordinate system}\lb{tras}
Basically, all the literature on some of the satellite orbital perturbations is developed in an equatorial coordinate system; thus, we need to devise a strategy to convert the existing analytical formulas for the equatorial rates of change of the satellite's orbital elements into expressions valid for the ecliptical coordinate system adopted here. Such an approach will turn out to be quite useful for gaining valuable information about, e.g., the tidal perturbations (see Section~\ref{tides}).

To this aim, let us start by rotating  the normal unit vector $\boldsymbol{\hat{n}}\textcolor{black}{=\grf{\sin I\sin\Omega,~-\sin I\cos\Omega,~\cos I}}$ \textcolor{black}{orthogonal to the orbital plane}, written in terms of the ecliptical elements, from the ecliptical to the equatorial system \textcolor{black}{by means of the rotation matrix}
\eqi
\textcolor{black}{\mathbb{R} = \left(
  \begin{array}{ccc}
    1 & 0 & 0 \\
    0 & \cos\epsilon & -\sin\epsilon \\
    0 & \sin\epsilon & \cos\epsilon \\
  \end{array}
\right)
,}
\eqf \textcolor{black}{where $\epsilon$ is the obliquity. T}he result is
\eqi
{\boldsymbol{\hat{n}}}^\textrm{eq} = \grf{\sin I \sin \Omega,~-\cos\epsilon \cos\Omega \sin I -\cos I \sin\epsilon,~\cos I \cos\epsilon - \cos\Omega \sin I \sin\epsilon }.\lb{un}
\eqf
Then,  let us calculate the node and the inclination referred to the equator from the components of \rfr{un} as
\begin{align}
I\eqt\ton{I,~\Omega;~\epsilon} \lb{Ieq}& = \arctan\ton{\rp{\sqrt{  \ton{{\hat{n}}^\textrm{eq}_x}^2 + \ton{{\hat{n}}^\textrm{eq}_y}^2   }}{ {\hat{n}}^\textrm{eq}_z  }}, \\ \nonumber\\
\Omega\eqt\ton{I,~\Omega;~\epsilon} \lb{Oeq}& = \arctan\ton{\rp{ {\hat{n}}^\textrm{eq}_x }{-{\hat{n}}^\textrm{eq}_y}}.
\end{align}
From \rfrs{Ieq}{Oeq} it turns out that $I,~\Omega = 90\deg$ correspond just to $I\eqt,~\Omega\eqt = 90\deg$.
By taking the time derivatives of \rfrs{Ieq}{Oeq}, it is possible to obtain exact analytical expressions of the rates of change of $I\eqt,~\Omega\eqt$ expressed in terms of their ecliptical $I,~\Omega$ counterparts. They get simplified for $I = \Omega = 90\deg$ reducing to\footnote{They hold also by accounting for $\dot\epsilon$.}
\begin{align}
\dot I\eqt & = \cos\epsilon~\dot I -\sin\epsilon~\dot\Omega, \\ \nonumber\\
\dot \Omega\eqt & = \sin\epsilon~\dot I + \cos\epsilon~\dot\Omega.
\end{align}
It is assumed that all the rates of changes appearing here and in the rest of the paper are averaged over the orbital period of the Earth's satellite and, when is the case, also over the period of an external third body;  for the sake of simplicity, the angular brackets $\ang{\ldots}$ denoting the average are omitted.
By solving with respect to the ecliptical rates of change $\dot I,~\dot\Omega$, one finally gets
\begin{align}
\dot I \lb{dotiec}& = \cos\epsilon~\dot I\eqt +\sin\epsilon~\dot\Omega\eqt, \\ \nonumber\\
\dot \Omega \lb{dotOec}& = -\sin\epsilon~\dot I\eqt + \cos\epsilon~\dot\Omega\eqt.
\end{align}
At this stage, there is nothing left to do but to express the known formulas for $\dot I\eqt,~\dot\Omega\eqt$ in terms of the ecliptical elements $I,~\Omega$. To this aim, it is useful to calculate $\cos I\eqt,~\sin I\eqt,~\cos\Omega\eqt,~\sin\Omega\eqt$ entering, e.g., the amplitudes of the tidal orbital perturbations. We have
\begin{align}
\cos I\eqt \lb{cosieq}& = {\bds{\hat{S}}}\eqt\bds\cdot{\boldsymbol{\hat{n}}}^\textrm{eq} =\cos I \cos\epsilon - \cos\Omega \sin I \sin\epsilon , \\ \nonumber \\
\sin^2 I\eqt \lb{sinieq}& = \left|{\bds{\hat{S}}}\eqt\bds\times{\boldsymbol{\hat{n}}}^\textrm{eq}\right|^2 = \ton{\cos\epsilon \cos\Omega \sin I + \cos I \sin\epsilon}^2 + \sin^2 I \sin^2\Omega,\\ \nonumber \\
\cos\Omega\eqt \lb{cosOeq} & =\rp{\cos\epsilon \cos\Omega \sin I +
 \cos I \sin\epsilon}{\sin I\eqt}, \\ \nonumber \\
\sin\Omega\eqt \lb{sinOeq} & =\rp{\sin I \sin\Omega}{\sin I\eqt}
\end{align}
\textcolor{black}{In \rfr{cosieq} and \rfr{sinieq}, ${\bds{\hat{S}}}\eqt$ is the unit vector of the spin axis of the rotating primary referred to its equator.}
%
\section{The Newtonian and post-Newtonian orbital rates of change}\lb{prece}
In the following, a circular orbit \textcolor{black}{with eccentricity $e = 0$} will be considered.
\subsection{The post-Newtonian Lense-Thirring effect}\lb{LTrate}
The long-term Lense-Thirring rates of change of the inclination and the node valid in any coordinate system in which the $\kx$ component of the primary's symmetry axis vanishes are \citep{2011PhRvD..84l4001I}
\begin{align}
\dot I_\textrm{LT} &= \rp{2GS\ky\sin\Omega}{c^2 a^3},\lb{LT}\\ \nonumber \\
\dot\Omega_\textrm{LT} & = \rp{2GS\ton{\kz + \ky\cot I\cos\Omega}}{c^2 a^3}\textcolor{black}{.}\lb{LTO}
\end{align}
\textcolor{black}{In \rfr{LT} and \rfr{LTO}, $G,~c$ are the Newtonian constant of gravitation and the speed of light in vacuum, $S$ is the primary's spin angular momentum, $a$ is the semimajor axis of the test particle's orbit.}
It should be noted that \rfr{LT} \textcolor{black}{and} \rfr{LTO} along with the following effects due to the geopotential (see \rfrs{OJ2}{IJ5} below) are just a particular case of general expressions valid in a completely arbitrary coordinate system in which \textcolor{black}{$\bds{\hat{S}}$} can assume any orientation in space \citep{2011PhRvD..84l4001I,2013JApA...34..341R,2014Ap&SS.352..493R}. If an ecliptical coordinate system is adopted, it can be demonstrated that the approach outlined in Section~\ref{tras} yields the same results as those obtained by \citet{2011PhRvD..84l4001I,2013JApA...34..341R,2014Ap&SS.352..493R} for $\kx=0,~\ky=\sin\epsilon,~\kz=\cos\epsilon$.

\begin{figure*}
\centerline{
\vbox{
\begin{tabular}{cc}
\epsfysize= 5.0 cm\epsfbox{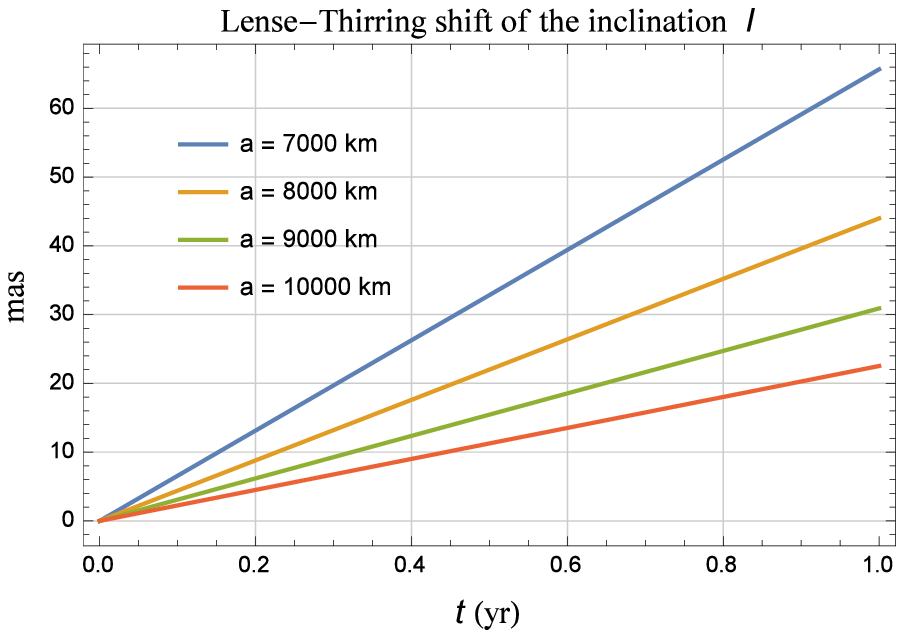} & \epsfysize= 5.0 cm\epsfbox{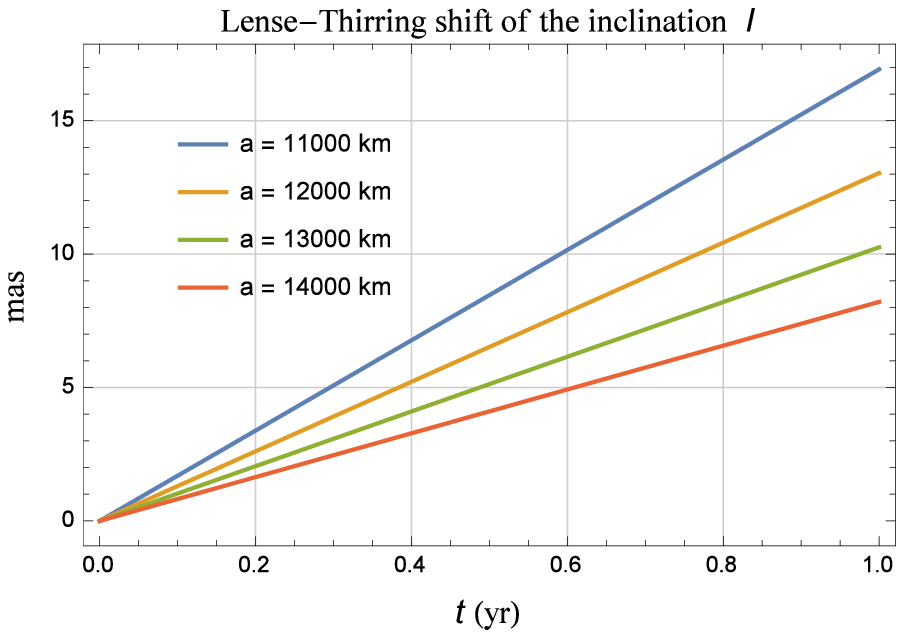}\\
\epsfysize= 5.0 cm\epsfbox{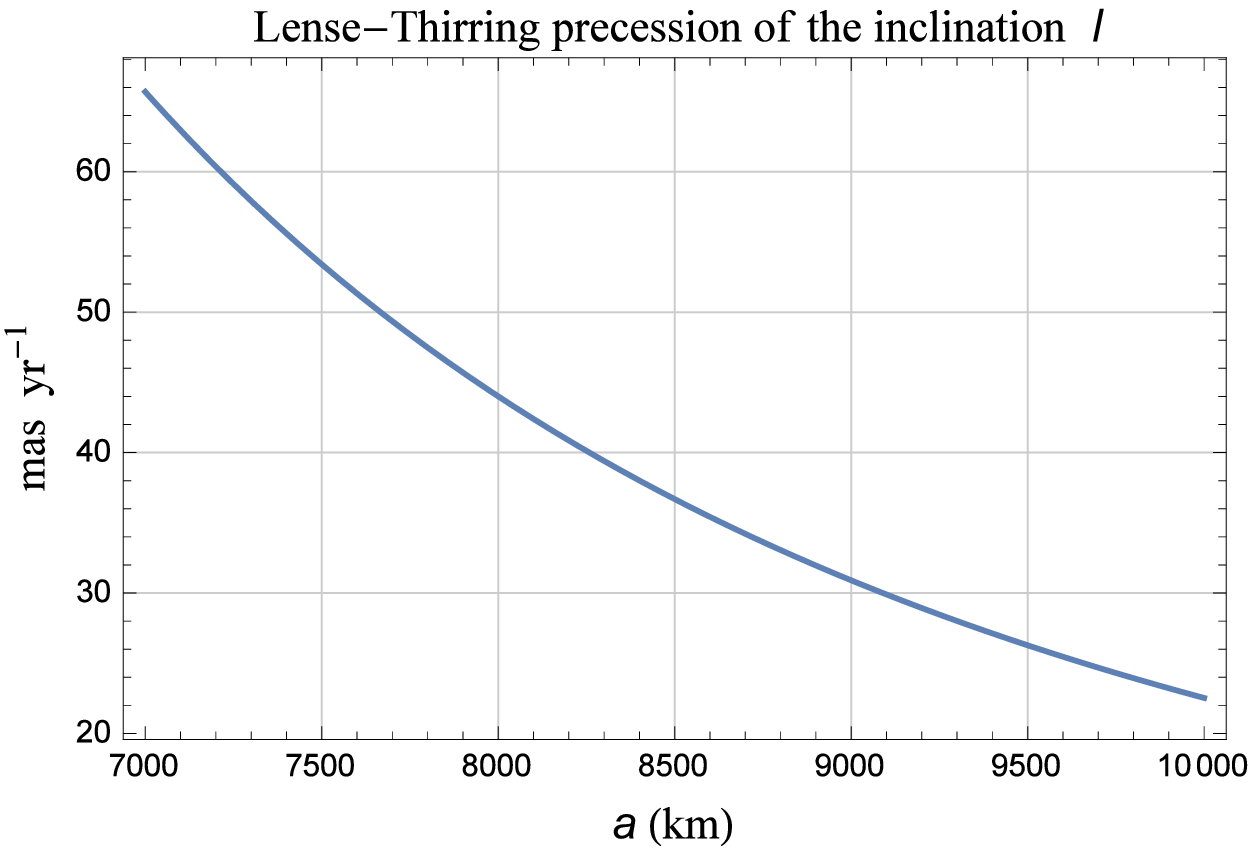} & \epsfysize= 5.0 cm\epsfbox{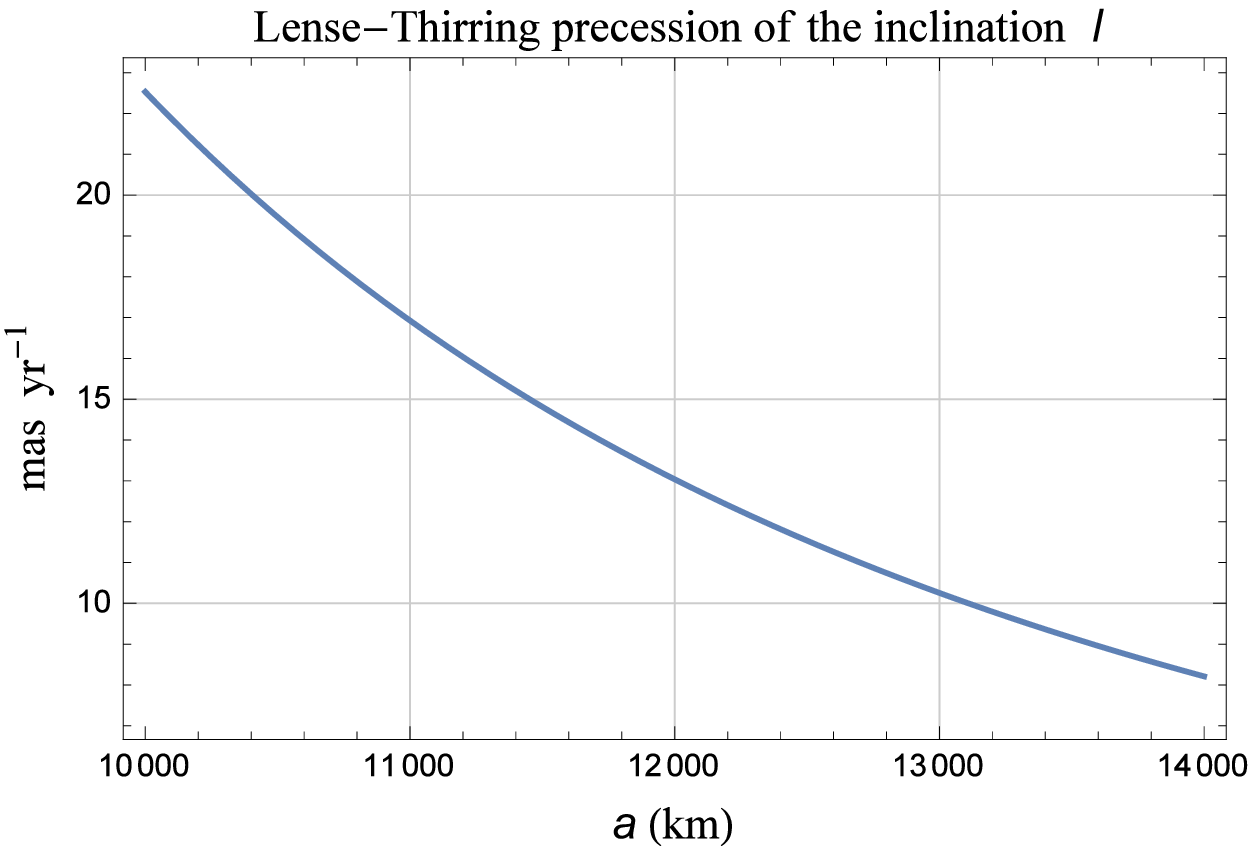}\\
\end{tabular}
}
}
\caption{Upper row: annual Lense-Thirring shifts of the satellite's inclination $I$ obtained for different values of $a$ by subtracting two time series produced by numerically integrating the equations of motion in rectangular Cartesian coordinates with and without the gravitomagnetic acceleration. Both the runs shared the same initial conditions characterized, among other things, by $e = 0,\Omega = I = 90\deg$. Lower row: Plot of the Lense-Thirring rate of change of the satellite's inclination $I$ as a function of the semimajor axis $a$ calculated analytically from \rfr{LT} for $e = 0,\Omega = I = 90\deg$. In both \textcolor{black}{cases}, a reference frame with the mean ecliptic at the epoch J2000.0 was used as reference $\grf{x,~y}$ plane so that $\kx=0,~\ky=\sin\epsilon = 0.3978,~\kz=\cos\epsilon = 0.9175$.}\label{fig1}
\end{figure*}

\begin{figure*}
\centerline{
\vbox{
\begin{tabular}{cc}
\epsfysize= 5.0 cm\epsfbox{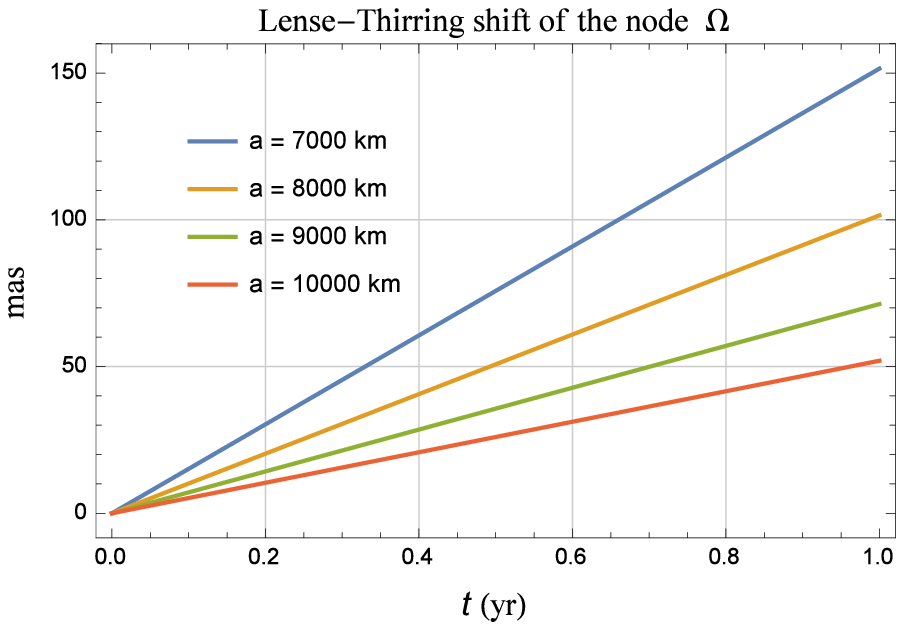} & \epsfysize= 5.0 cm\epsfbox{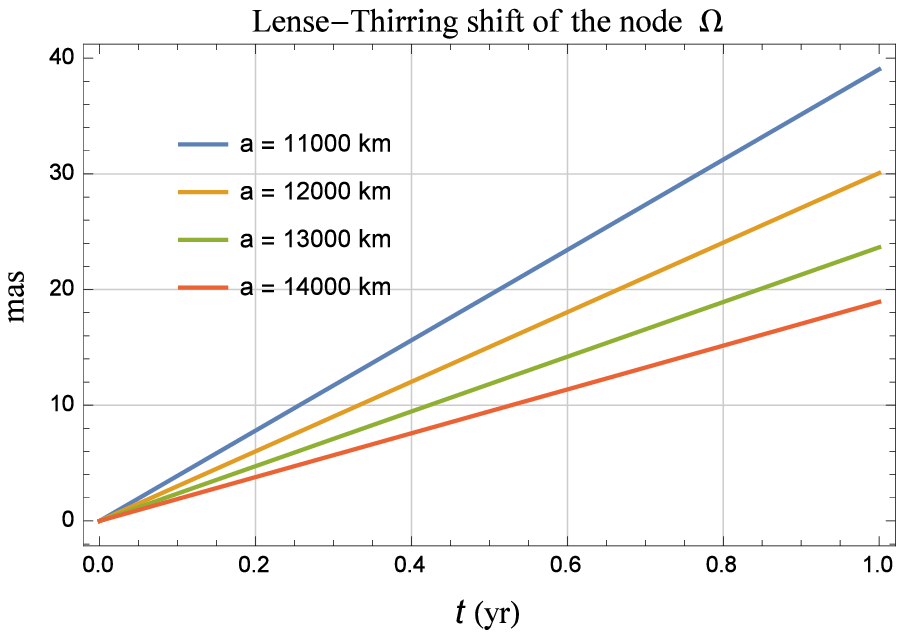}\\
\epsfysize= 5.0 cm\epsfbox{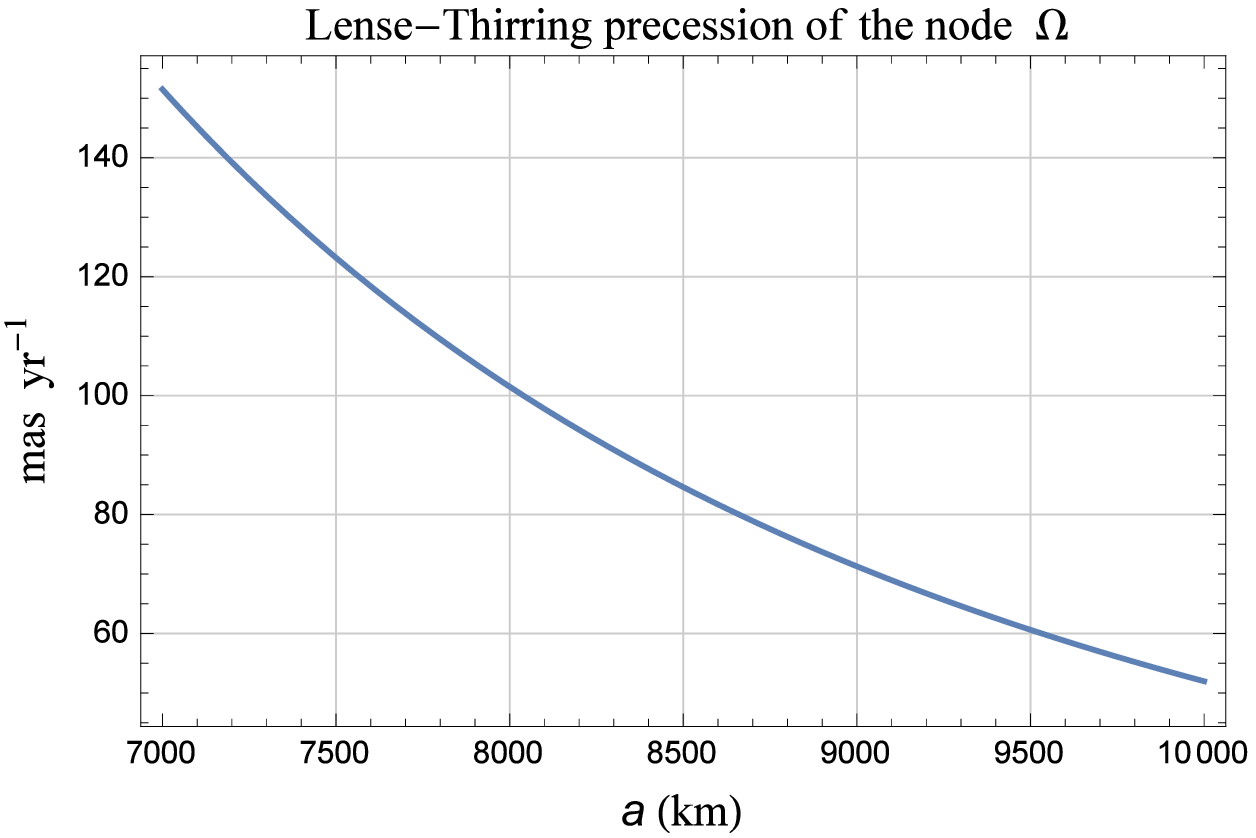} & \epsfysize= 5.0 cm\epsfbox{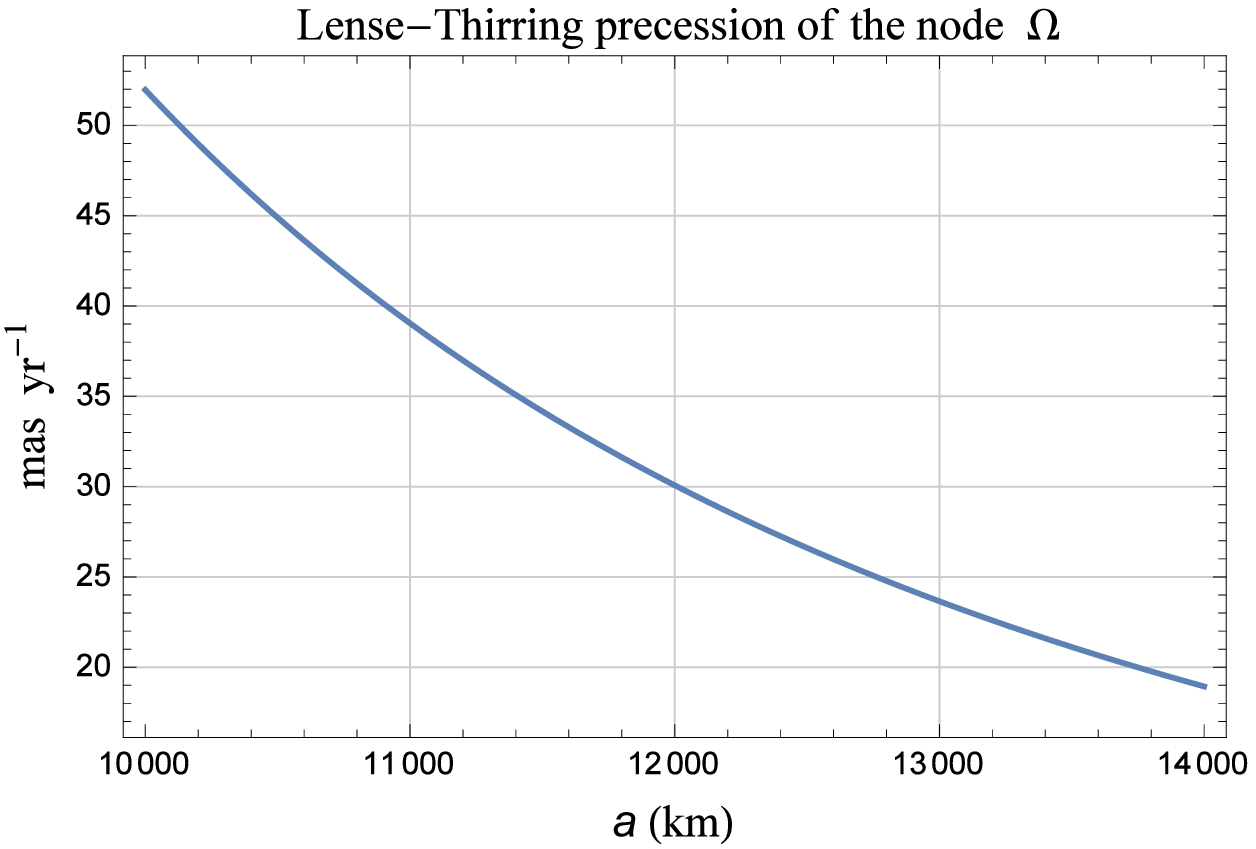}\\
\end{tabular}
}
}
\caption{Upper row: annual Lense-Thirring shifts of the satellite's node $\Omega$ obtained for different values of $a$ by subtracting two time series produced by numerically integrating the equations of motion in rectangular Cartesian coordinates with and without the gravitomagnetic acceleration. Both the runs shared the same initial conditions characterized, among other things, by $e = 0,\Omega = I = 90\deg$. Lower row: Plot of the Lense-Thirring rate of change of the satellite's node $\Omega$ as a function of the semimajor axis $a$ calculated analytically from \rfr{LTO} for $e = 0,\Omega = I = 90\deg$. In both \textcolor{black}{cases}, a reference frame with the mean ecliptic at the epoch J2000.0 was used as reference $\grf{x,~y}$ plane so that $\kx=0,~\ky=\sin\epsilon = 0.3978,~\kz=\cos\epsilon = 0.9175$.}\label{fig2}
\end{figure*}

Figure~\ref{fig1} \textcolor{black}{and} Figure~\ref{fig2} depict the agreement between \rfr{LT} \textcolor{black}{and} \rfr{LTO} and the numerically integrated Lense-Thirring shifts which display just the expected linear temporal behaviour for the specific scenario $I=\Omega=90\deg$ (see Section~\ref{zonrate} for its relevance).
\subsection{The Newtonian even and odd zonal harmonics of the geopontial}\lb{zonrate}
The classical long-term rates of change of the node due to the first even \textcolor{black}{and odd} zonals \textcolor{black}{of low degree} are\footnote{Eqs.~(12)~to~(15) of \citet{2011PhRvD..84l4001I} yield \rfr{IJ2} and \rfr{OJ2} with the replacement $Q_2\rightarrow -GM R^2 J_2$. It corrects a missing minus sign in \citet[p. 124001-4]{2011PhRvD..84l4001I}.} \citep{2011PhRvD..84l4001I,2013JApA...34..341R,2014Ap&SS.352..493R}
\begin{align}
\dot\Omega_{J_2} &= \rp{3}{2}\nk J_2 \ton{\rp{R}{a}}^2 \ton{\ky \cos\Omega - \kz \cot I}\ton{\kz + \ky \cos\Omega \cot I}\sin I,\lb{OJ2} \\ \nonumber \\
\dot\Omega_{J_3} & = 0, \\ \nonumber \\
\dot\Omega_{J_4} \nonumber &= -\rp{15}{64} \nk J_4 \ton{\rp{R}{a}}^4\ton{\kz + \ky \cos \Omega \cot I}\ton{\kz \cos  I - \ky \cos \Omega \sin  I}\times\\ \nonumber \\
&\times \qua{5 - 7 \kz^2 +\ton{7 - 21 \kz^2}\cos 2I + 14 \ton{-1 + \kz^2} \cos 2\Omega \sin^2 I + 28 \ky \kz \cos \Omega \sin 2I},\lb{OJ4} \\ \nonumber \\
\textcolor{black}{\dot\Omega_{J_5}}  & \textcolor{black}{= 0.}
\end{align}

The classical long-term rates of change of the inclination due to the first even and odd zonals of low degree are \citep{2011PhRvD..84l4001I,2013JApA...34..341R,2014Ap&SS.352..493R}
\begin{align}
\dot I_{J_2} \lb{IJ2}&= \rp{3}{2}\nk J_2 \ton{\rp{R}{a}}^2\ky\ton{\ky\sin I\cos\Omega - \kz\cos I}\sin\Omega, \\ \nonumber \\
\dot I_{J_3} &= 0\lb{IJ3}, \\ \nonumber \\
\dot I_{J_4} \nonumber &= \rp{15}{128}\nk J_4 \ton{\rp{R}{a}}^4\ky\grf{-\ky \ton{-1 + 7 \kz^2}\ton{5 + 7 \cos 2I} \sin I \sin 2 \Omega + \right.\\ \nonumber \\
\nonumber &+\left. \kz \cos I \qua{\ton{-3 + 7 \kz^2}\ton{1 + 7 \cos 2I} \sin \Omega -42 \ton{-1 + \kz^2} \sin^2 I \sin 3 \Omega} + \right.\\ \nonumber \\
&+\left. 7 \ky \ton{-1 + \kz^2} \sin^3 I\sin 4 \Omega },\lb{IJ4} \\ \nonumber \\
\dot I_{J_5} \lb{IJ5}&= 0.
\end{align}
\textcolor{black}{In \rfrs{OJ2}{IJ5}, $\nk=\sqrt{\mu a^{-3}}$ is the Keplerian mean motion ($\mu=GM$ is the gravitational parameter of the primary, whose mass is $M$), $R$ is the mean equatorial radius of the central body, while $J_\ell = -\sqrt{2\ell + 1}~{\overline{C}}_{\ell,0},~\ell=2,3,4,\ldots$ are the zonal harmonic coefficients of degree $\ell$  of the Newtonian multipolar expansion of the primary's gravity field. The latter ones, in turn, are expressed in terms of  ${\overline{C}}_{\ell,m}$, which are the fully normalized Stokes coefficient of degree $\ell$ and order $m$ of the multipolar expansion of the gravitational potential of the central body; $m=0$ for the zonal harmonics.}

For a general value of the inclination, \rfrs{OJ2}{OJ4} tells us that, if $\Omega=90\deg$, the node circulates with a secular precession given mainly by
\eqi
\dot\Omega \simeq -\rp{3}{2}\nk J_2 \ton{\rp{R}{a}}^2\kz^2\cos I\qua{1  -  \rp{5}{8}\rp{J_4}{J_2}\ton{\rp{R}{a}}^2 \ton{-3 + 7 \kz^2 \cos^2 I}}.\lb{OJ290}
\eqf
This implies that the inclination undergoes both relativistic and classical long-periodic, harmonic variations whose frequencies are  $j\dot\Omega,~j=1,2,3\ldots$. In particular, there are some components of \rfrs{IJ2}{IJ5}, proportional to $\ky\kz\cos I\sin\Omega$,  which have the same temporal pattern of \rfr{LT}. Thus, they act as a potentially insidious systematic bias depending on the level of mismodeling in the zonal harmonics. The same holds also for \rfr{LTO}, impacted by the mismodelled part of \rfr{OJ290}.

\begin{figure*}
\centerline{
\vbox{
\begin{tabular}{cc}
\epsfysize= 5.0 cm\epsfbox{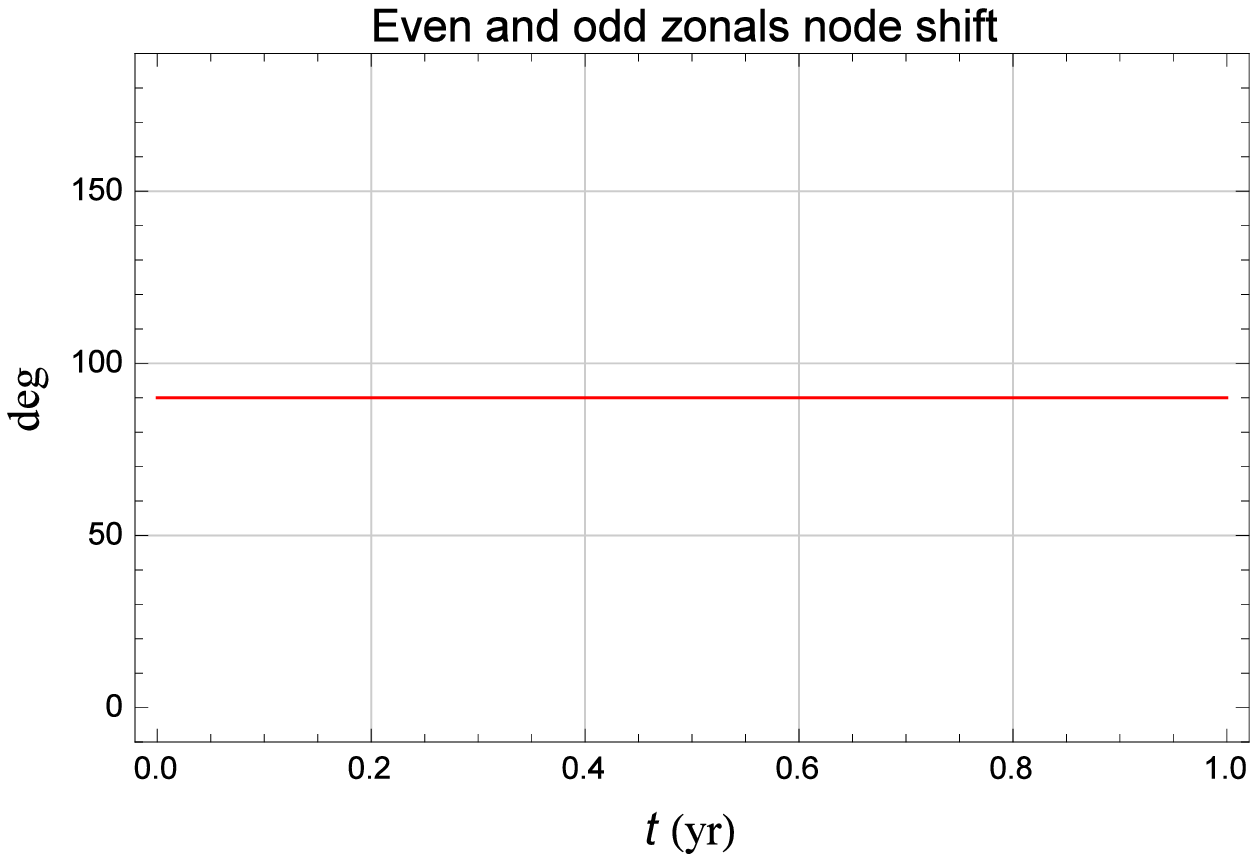} & \epsfysize= 5.0 cm\epsfbox{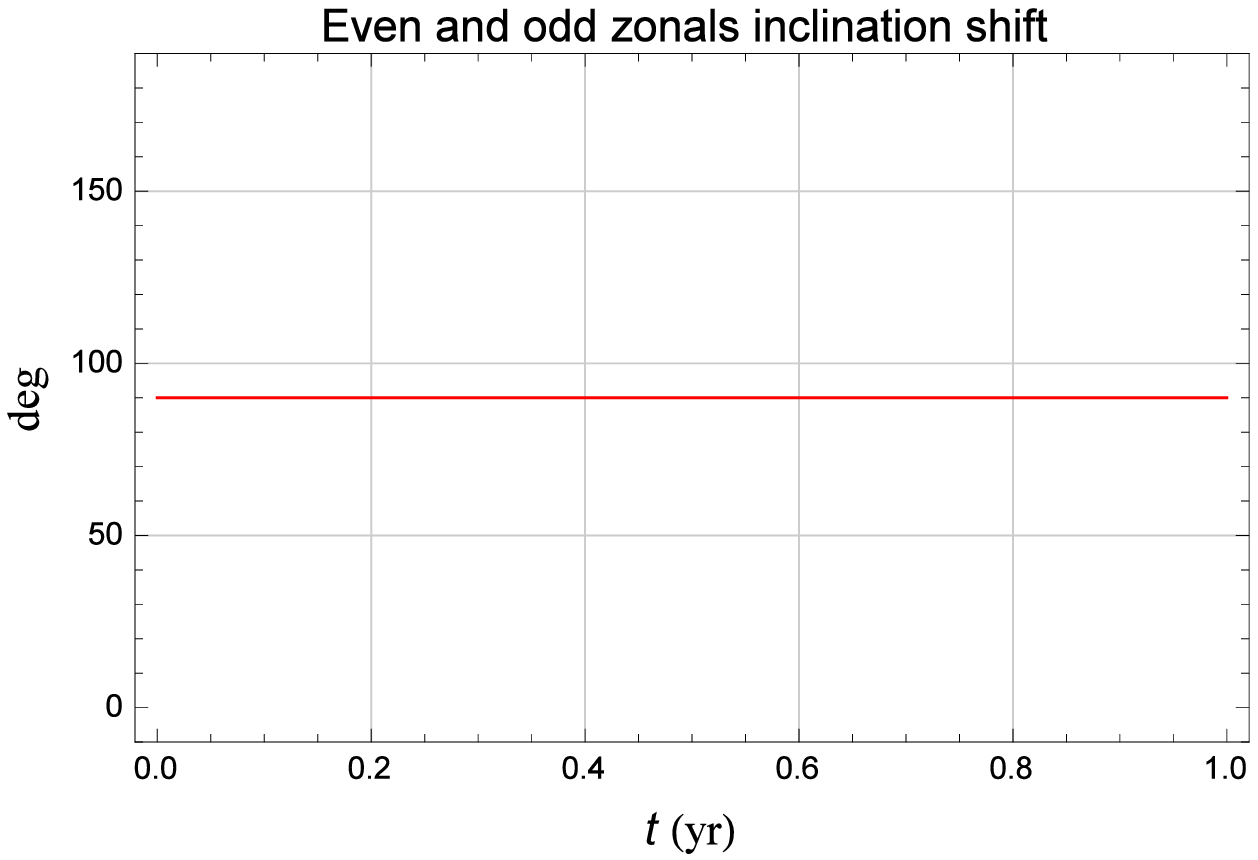}\\
\end{tabular}
}
}
\caption{Annual shifts of the satellite's node $\Omega$ (left panel) and inclination $I$ (right panel) obtained, for each orbital element, by subtracting two time series produced by numerically integrating the equations of motion in rectangular Cartesian coordinates with and without the classical accelerations due to the first five  zonal harmonics $J_2,~J_3,~J_4,~J_5,~J_6$ of the geopotential. For each orbital element, both the runs shared the same initial conditions characterized, among other things, by $e = 0,\Omega = I = 90\deg$. The meaning of the plots displayed is that if the satellite's node and inclination are set to such initial values, they stay constant to them throughout the orbital evolution. The result turns out to be independent of the semimajor axis $a$. A reference frame with the mean ecliptic at the epoch J2000.0 was used as reference $\grf{x,~y}$ plane so that $\kx=0,~\ky=\sin\epsilon = 0.3978,~\kz=\cos\epsilon = 0.9175$.}\label{fig3}
\end{figure*}

On the other hand, if the orbital plane is perpendicular to the ecliptic ($I = 90\deg$), by choosing the initial value $\Omega = 90\deg$ allows to:
\begin{enumerate}[(a)]
\item Keep the node \textcolor{black}{rate} essentially constant, as per
\begin{align}
\dot\Omega_{J_2} &= \rp{3}{2}\nk J_2 \ton{\rp{R}{a}}^2\kz\ky\cos\Omega,\lb{OJ2p}\\ \nonumber \\
\dot\Omega_{J_4} &= \rp{15}{32}\nk J_4 \ton{\rp{R}{a}}^4\kz\ky\qua{-1 + 7\kz^2 +7\ton{-1+\kz^2}\cos 2\Omega}\cos\Omega\lb{OJ4p}
\end{align}
which are obtained from \rfrs{OJ2}{OJ4} for $I=90\deg$.
\item Maximize the Lense-Thirring rates of change which \textcolor{black}{become}  secular trends, as per \rfr{LT} \textcolor{black}{and} \rfr{LTO}
\item  Cancel all the classical rates of change on the inclination due to the static part of the geopotential, as per
\begin{align}
\dot I_{J_2} \lb{IJ2p}& = \rp{3}{4}\nk\ton{\rp{R}{a}}^2{\ky}^2 J_2\sin 2\Omega, \\ \nonumber \\
\dot I_{J_3} \lb{IJ3p}& = 0, \\ \nonumber \\
\dot I_{J_4} \lb{IJ4p}& = \rp{15}{64}\nk\ton{\rp{R}{a}}^4 J_4\ton{1-\kz^2}\qua{-1 + 7\kz^2 +7\ton{-1+\kz^2}\cos 2\Omega}\sin 2\Omega, \\ \nonumber \\
\dot I_{J_5} \lb{IJ5p}& = 0,
\end{align}
which come from \rfrs{IJ2}{IJ5} for $I=90\deg$.
\end{enumerate}
Figure~\ref{fig3}, obtained by numerically integrating the equations of motion with the accelerations due to the first five zonals of the geopotential, shows that, actually, the node and the inclination stay constant to their initial values if $\Omega = I = 90\deg$ are adopted for them.

\begin{figure*}
\centerline{
\vbox{
\begin{tabular}{cc}
\epsfysize= 5.0 cm\epsfbox{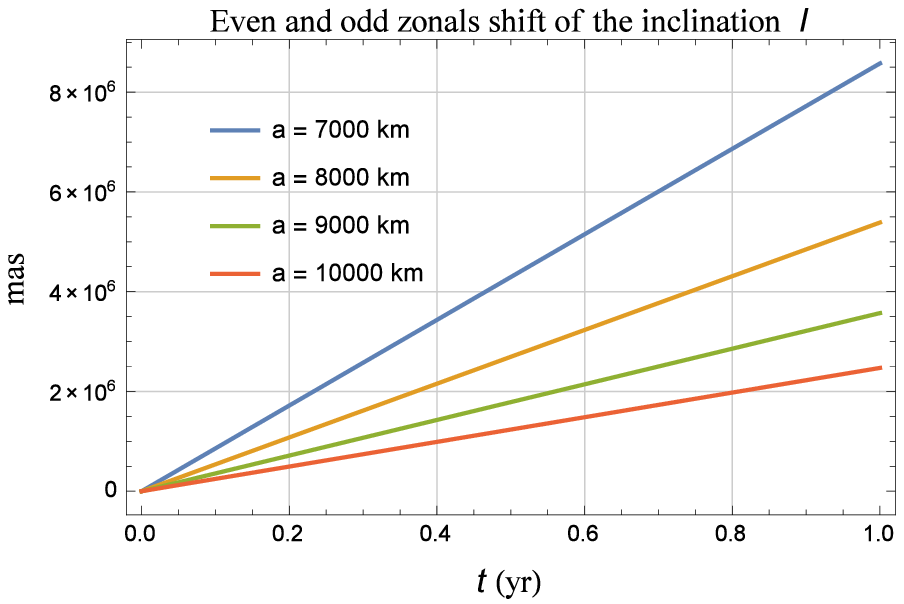} & \epsfysize= 5.0 cm\epsfbox{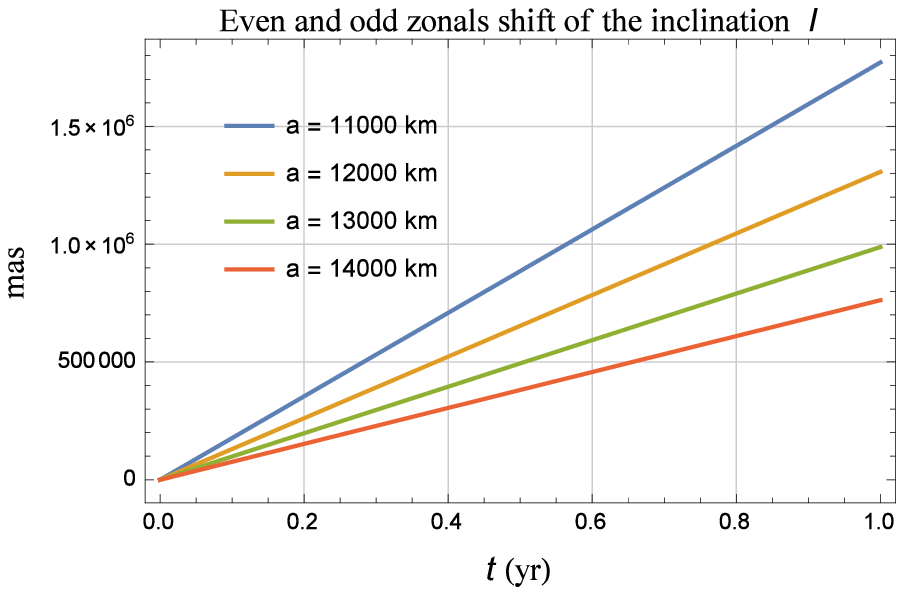}\\
\epsfysize= 5.0 cm\epsfbox{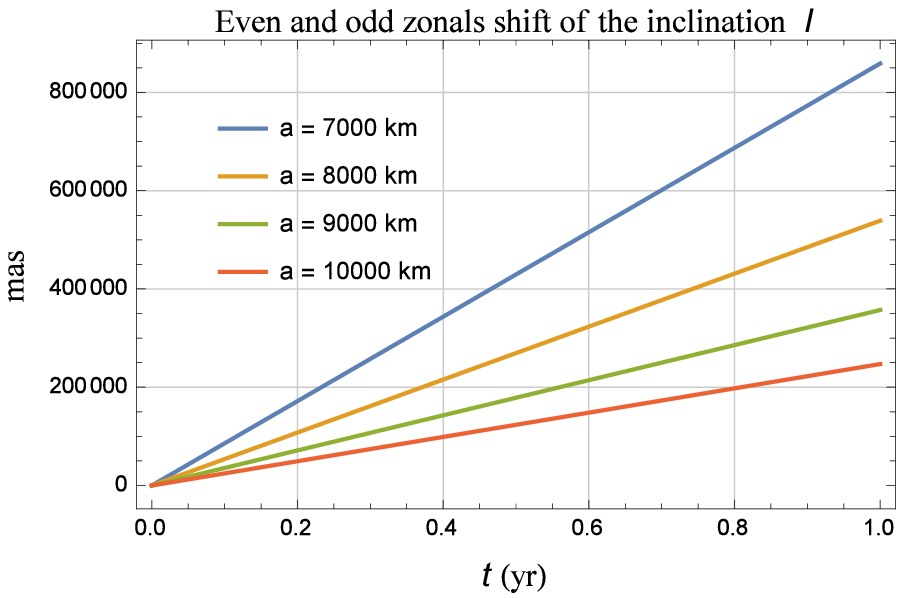} & \epsfysize= 5.0 cm\epsfbox{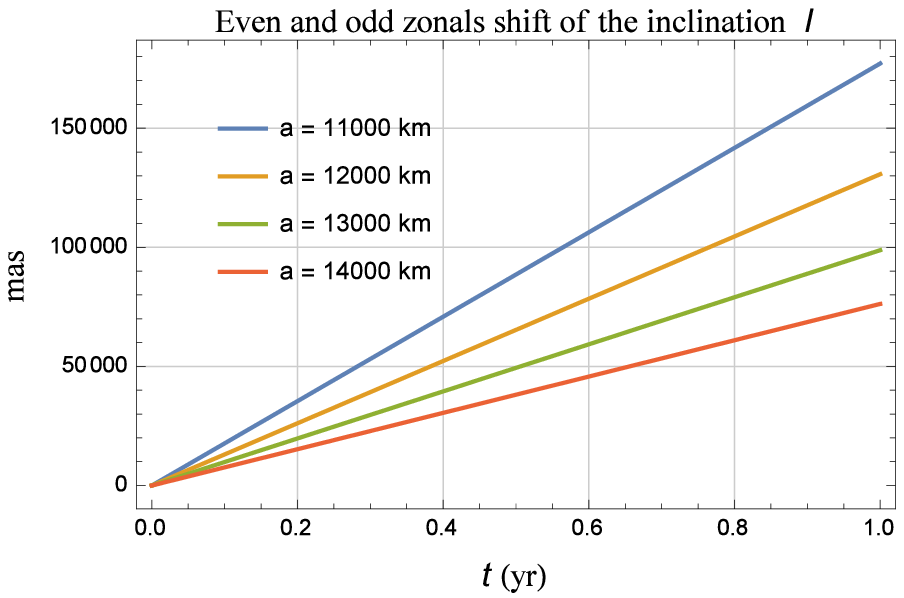}\\
\end{tabular}
}
}
\caption{Upper row: nominal annual shifts of the satellite's  inclination $I$  obtained for different values of $a$ by subtracting two time series produced by numerically integrating the equations of motion in rectangular Cartesian coordinates with and without the classical accelerations due to the first five  zonal harmonics $J_2,~J_3,~J_4,~J_5,~J_6$ of the geopotential. Both the runs shared the same initial conditions characterized, among other things, by $e = 0,\Omega = I = 90\pm 0.1\deg$. Lower row: same as in the upper row, apart from on offset of $0.01\deg$ from the ideal condition $I=\Omega=90\deg$. The largest contribution is due to $J_2$, whose present-day uncertainty may be as large as $\lesssim 2\times 10^{-10}$ if evaluated conservatively; the statistical, formal errors $\upsigma_{{\overline{C}}_{2,0}}$ released in the global gravity field models produced from the GRACE/GOCE data by several institutions around the world are even $\simeq 1-3$ orders of magnitude smaller. In both cases, a reference frame with the mean ecliptic at the epoch J2000.0 was used as reference $\grf{x,~y}$ plane so that $\kx=0,~\ky=\sin\epsilon = 0.3978,~\kz=\cos\epsilon = 0.9175$.}\label{fig4}
\end{figure*}

\begin{figure*}
\centerline{
\vbox{
\begin{tabular}{cc}
\epsfysize= 5.0 cm\epsfbox{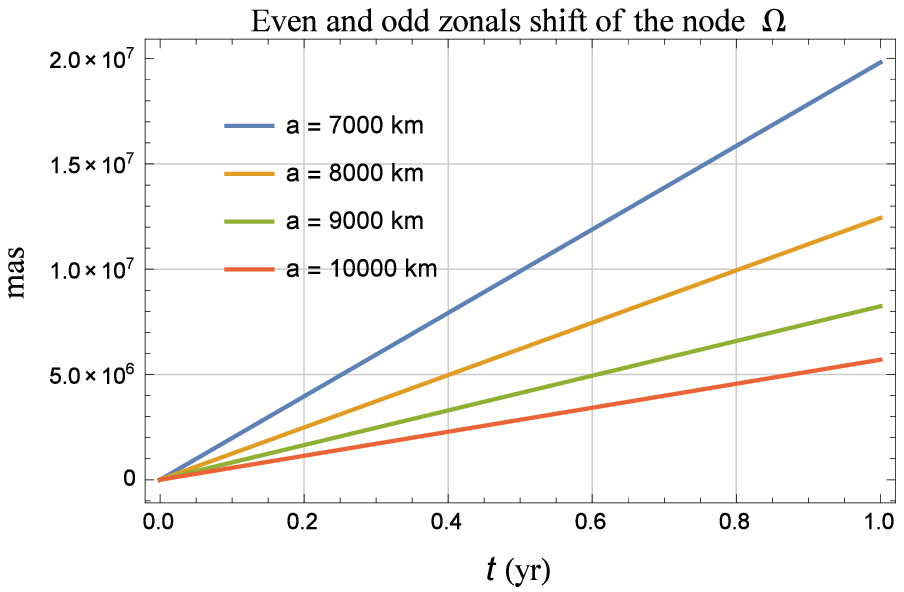} & \epsfysize= 5.0 cm\epsfbox{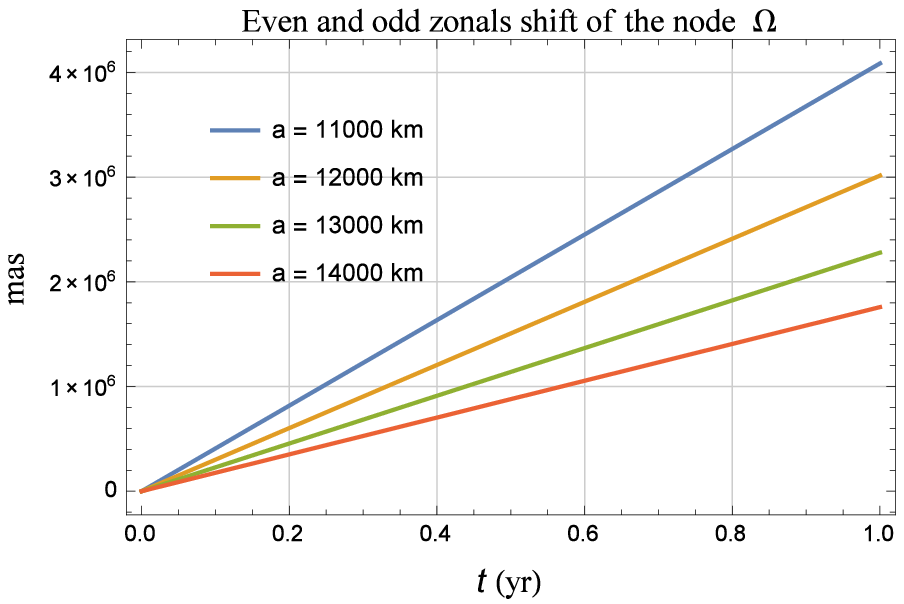}\\
\epsfysize= 5.0 cm\epsfbox{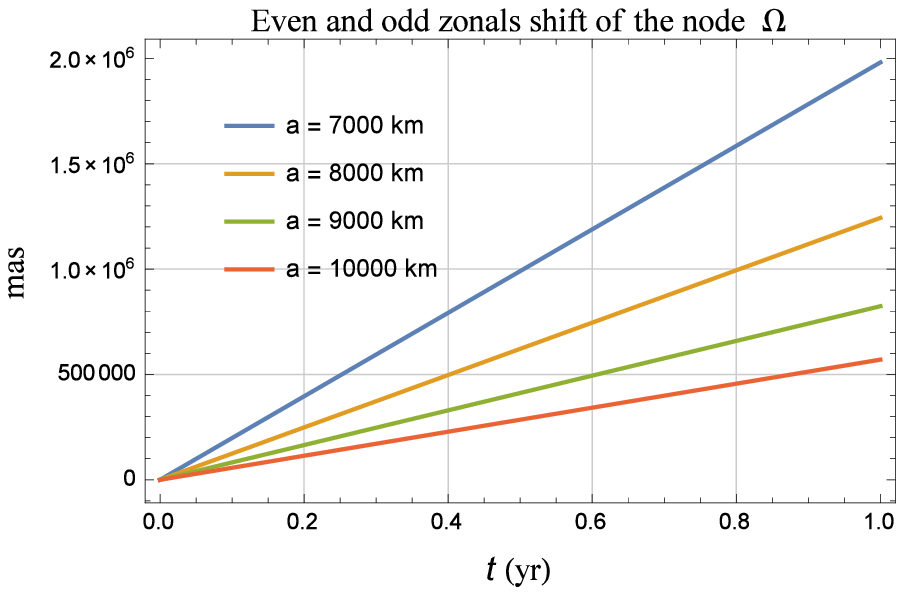} & \epsfysize= 5.0 cm\epsfbox{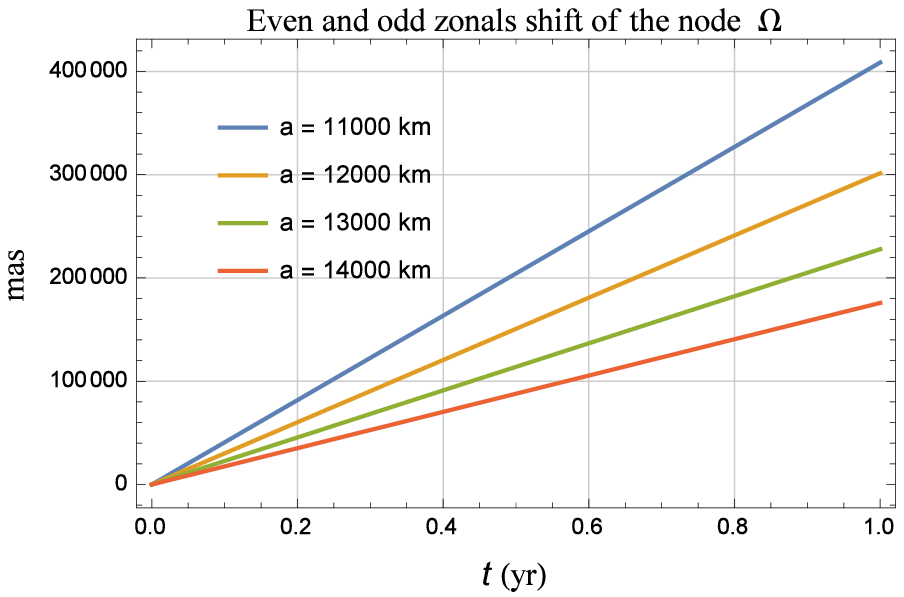}\\
\end{tabular}
}
}
\caption{Upper row: nominal annual shifts of the satellite's  node $\Omega$  obtained for different values of $a$ by subtracting two time series produced by numerically integrating the equations of motion in rectangular Cartesian coordinates with and without the classical accelerations due to the first five zonal harmonics $J_2,~J_3,~J_4,~J_5,~J_6$ of the geopotential. Both the runs shared the same initial conditions characterized, among other things, by $e = 0,\Omega = I = 90\pm 0.1\deg$. Lower row: same as in the upper row, apart from on offset of $0.01\deg$ from the ideal condition $I=\Omega=90\deg$. The largest contribution is due to $J_2$, whose present-day uncertainty may be as large as $\lesssim 2\times 10^{-10}$ if evaluated conservatively; the statistical, formal errors $\upsigma_{{\overline{C}}_{2,0}}$ released in the global gravity field models produced from the GRACE/GOCE data by several institutions around the world are even $\simeq 1-3$ orders of magnitude smaller. In both cases, a reference frame with the mean ecliptic at the epoch J2000.0 was used as reference $\grf{x,~y}$ plane so that $\kx=0,~\ky=\sin\epsilon = 0.3978,~\kz=\cos\epsilon = 0.9175$.}\label{fig5}
\end{figure*}

In order to cope with the unavoidable orbital injection errors inducing departures from the chosen ideal orbital geometry, in Figure~\ref{fig4} \textcolor{black}{and} Figure~\ref{fig5} we numerically investigate the impact of offsets  of the order of $\Delta\Omega = \Delta I = 0.1-0.01\deg$ from the proposed scenario characterized by $\Omega = I = 90\deg$ for different altitudes of the satellite. Since the largest contribution to the classical inclination rate is due to $J_2$, the level of mismodeling in it plays a crucial role in determining the largest admissible deviations from the nominal orbital configuration. According to \citet{2012JHEP...05..073I}, who relies upon the method proposed by \citet{2012JGeod..86...99W} to realistically compare geopotential harmonics in recent and past gravitational fields, a conservative evaluation of the actual uncertainty in the first even zonal points toward  $\delta{\overline{C}}_{2,0}\simeq 3\times 10^{-11}-2\times 10^{-10}$.
On the other hand, the formal, statistical errors $\upsigma_{{\overline{C}}_{2,0}}$ released in several global gravity models are as little as $\simeq 10^{-12}-10^{-13}$.
By assuming $\delta{\overline{C}}_{2,0} = 2\times 10^{-10}$, Figure~\ref{fig1} \textcolor{black}{and} Figure~\ref{fig2} and Figure~\ref{fig4} \textcolor{black}{and} Figure~\ref{fig5} show that $\Delta\Omega = \Delta I = 0.01\deg$  allow to reach a $\simeq 3-5\times 10^{-3}$ level of systematic error in the Lense-Thirring effect for any altitude considered, while for $\Delta\Omega,~\Delta I = 0.1\deg$, the bias amounts to $\simeq 3-5\times 10^{-2}$.
\textcolor{black}{Such results show that, in the present case and contrary to other ongoing and forthcoming tests of the gravitomagnetic field of the Earth with, e.g., the LAGEOS type spacecraft, it is not so important to have a particularly accurate value of ${\overline{C}}_{2,0}$, at least with respect to such other contexts. Indeed, even should our choice for $\delta{\overline{C}}_{2,0}$ be too conservative, this could only further benefit our scenario allowing for even weaker constraints on the offsets  $\Delta I,~\Delta\Omega$. Thus, we can put aside certain subtleties pertaining, e.g., the different mean epochs of the geopotential models used, their time dependence, the role of the a-priori background gravity fields used in constructing the new global solutions, the different accuracies at the low degrees of the models based on GRACE and GOCE, etc. They may enter a more detailed discussion on the correct way to asses the realistic uncertainty in ${\overline{C}}_{2,0}$ which might be more pertinent in different tests.}
%
It is interesting to remark that the deviations from the ideal polar orbit of GP-B were as little as $5\times 10^{-5}\deg$ at its launch \citep[p. 141]{gpbrep}; our constraints are much less demanding.
%
%
%
%
%
%
%
%

\begin{figure*}
\centerline{
\vbox{
\begin{tabular}{cc}
\epsfysize= 5.0 cm\epsfbox{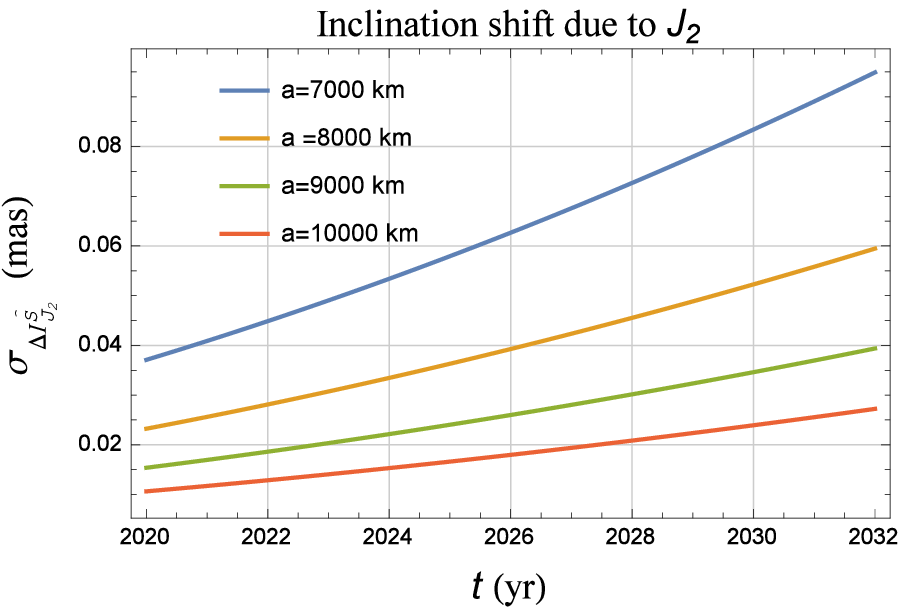} & \epsfysize= 5.0 cm\epsfbox{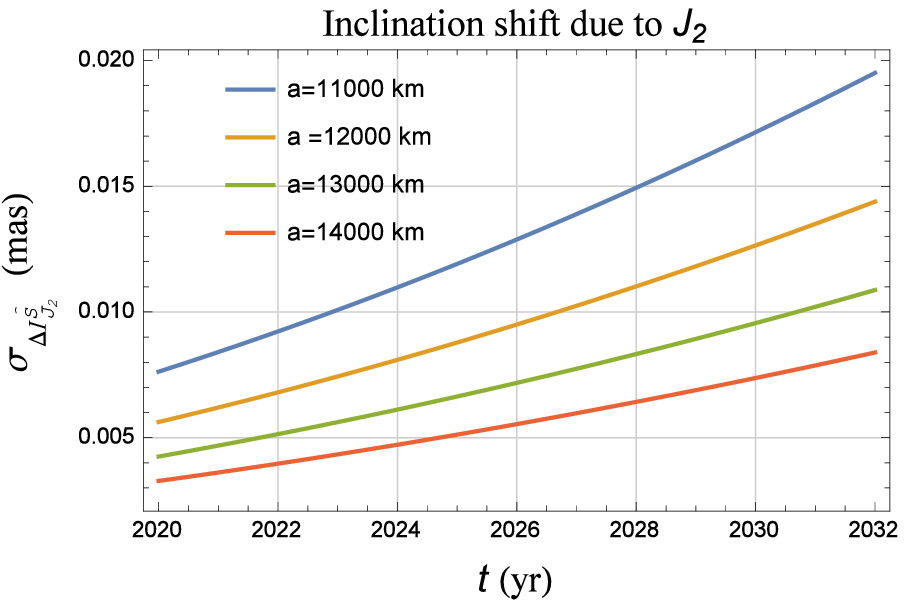}\\
\epsfysize= 5.0 cm\epsfbox{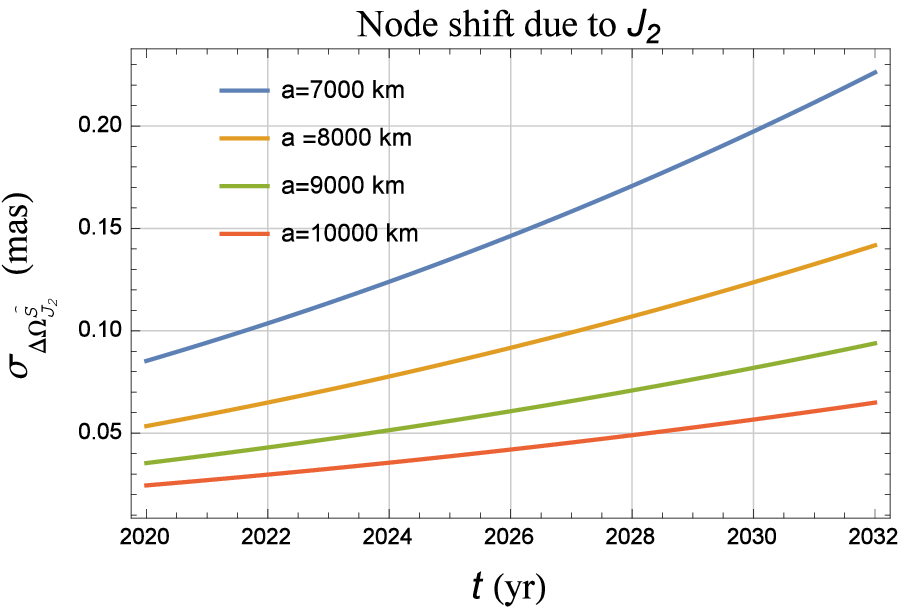} & \epsfysize= 5.0 cm\epsfbox{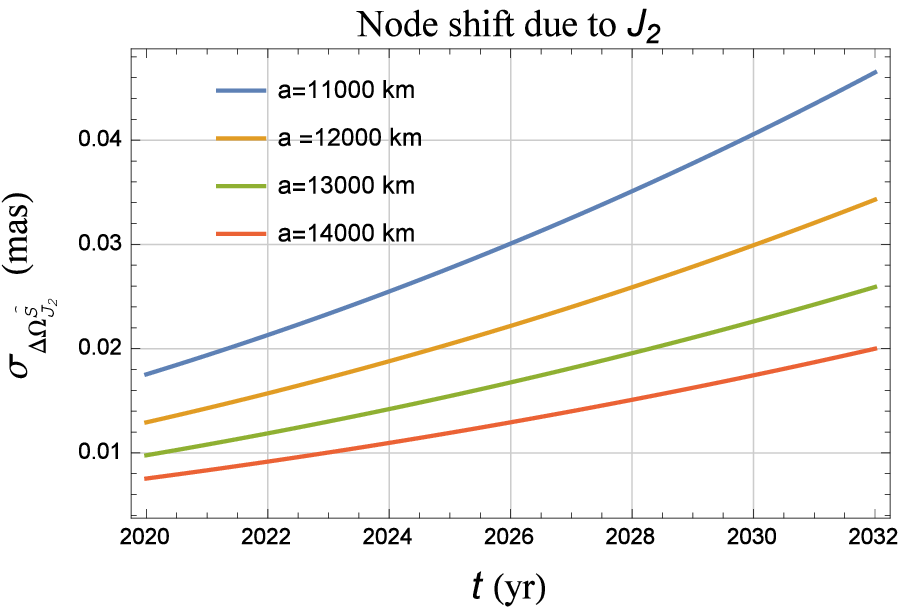}\\
\end{tabular}
}
}
\caption{\textcolor{black}{
Upper row: Mismodeled time series $\upsigma_{\Delta I_{J_2}^{\bds{\hat{S}}}}\ton{t}$, produced analytically from \citet[Eqs.~(12)~to~(15)]{2011PhRvD..84l4001I} for different values of the semimajor axis $a$ and $I_0=\Omega_0 = 90 \pm 0.01\deg$, of the time-dependent $J_2$-induced shift of the inclination $I$ due to the uncertainties of the parameters entering the precession/nutation and the temporal change of the obliquity according to the values listed in
Table~\ref{nutt} of Appendix~\ref{appenb}. The nominal value of $J_2$, retrieved from some model, was adopted. Lower row: same as for the node.
}}\label{figNUT}
\end{figure*}

\textcolor{black}{
In fact, the long-term geopotential perturbations bring indirectly into play another potential source of systematic error. It is the time-dependence of the Earth's spin axis because of the precession and nutation induced by the time-varying lunisolar torques, which displace $\bds{\hat{S}}$ from its orientation with respect to the mean equator and equinox of J2000.0 to the true equator and equinox (true-of-date), and the of the obliquity itself, which experiences a slight decrease, because of the gravitational pull of exerted by the other planets; see, e.g., \citet[Sect.~5.2]{2000Monte}. Such effects induce a non-linear time dependence on the inclination and node rates of change averaged over the satellite's orbital period which, in principle, should be taken into account in evaluating the temporal changes of $I,~\Omega$ mainly due to $J_2$ since, as a result, they undergo additional offsets $\Delta I_{J_2}^{\bds{\hat{S}}}\ton{t},~\Delta\Omega_{J_2}^{\bds{\hat{S}}}\ton{t}$ induced by the aforementioned astronomical phenomena. In particular, it is important that they can be modeled with sufficiently high precision for our accuracy goals. Since we are not interested here in, say, paleoclimatological studies spanning temporal intervals as long as Myr or Gyr, it can be done as follows. First, let us take the Earth's spin axis referred to the mean equator and equinox at J2000.0 ${\bds{\hat{S}}}_0=\grf{0,0,1}$ and  refer it to the mean equator and equinox of some other epoch $t$ (mean-of-date) by means of the precession matrix $\mathbb{P}$. As a suitable parametrisation of it, the three angles $\zeta,~\vartheta,~z$ can be adopted; over timescales of just some yr, they can be expressed as
}
\textcolor{black}{
\begin{align}
\zeta(t) & \simeq \zeta_0 + \dot\zeta t +\mathcal{O}\ton{t^2},\lb{zita} \\ \nonumber \\
\vartheta(t) & \simeq \dot\vartheta t +\mathcal{O}\ton{t^2}, \\ \nonumber \\
z(t) & \simeq z_0 + \dot z t +\mathcal{O}\ton{t^2}.\lb{zetta}
\end{align}
}
\textcolor{black}{
The values and the associated uncertainties of the quantities entering \rfrs{zita}{zetta} are listed in Table~\ref{nutt} of Appendix~\ref{appenb}.
Then, the mean-of-date spin axis has to be rotated to its true-of-date orientation by means of the nutation matrix $\mathbb{N}$ expressed in terms of the angles $\epsilon + \Delta\epsilon,~\epsilon,~\Delta\psi$. About $\Delta\epsilon,~\Delta\psi$, we will retain just their largest contributions due to the lunar node $\Omega_{\leftmoon}$, whose relevant values and uncertainties are displayed in Table~\ref{nutt} of Appendix~\ref{appenb}. Finally, we will rotate $\bds{\hat{S}}$ from the true-of-date coordinates, referred to the true equator and equinox, to the ecliptical coordinates by accounting for the rate of change of the obliquity as
}
\textcolor{black}{
\eqi
\epsilon\simeq \epsilon_0 + \dot\epsilon t + \mathcal{O}\ton{t^2};
\eqf
}
\textcolor{black}{
Table~\ref{nutt} of Appendix~\ref{appenb} collects the values and the uncertainties of $\epsilon_0,~\dot\epsilon$.
As a result, the exact long-term rates of change of $I,~\Omega$ due to $J_2$ written in terms of $\kx,~\ky,~\kz$, retrievable from \citet[Eqs.~(12)~to~(15)]{2011PhRvD..84l4001I}, fully account for the motion of the Earth's spin axis. The, we integrate the Taylor expansion of $\dot I_{J_2}\ton{t},~\dot\Omega_{J_2}\ton{t}$ truncated to, say, the 3rd order in $t$, with respect to time obtaining analytical expressions for the time-dependent offsets $\Delta I^{\bds{\hat{S}}}_{J_2}\ton{t},~\Delta\Omega^{\bds{\hat{S}}}_{J_2}\ton{t}$ due to the temporal evolution of ${\bds{\hat{S}}}\ton{t}$. Finally, we propagate the present-day errors in $\epsilon_0,~\dot\epsilon,~\zeta_0,~\dot\zeta,~\dot\vartheta,~z_0,~\dot z,~\Delta\epsilon,~\Delta\psi$ in a root-sum-square (RMS) way by obtaining the time series of the mismodeled offsets $\upsigma_{\Delta I^{\bds{\hat{S}}}_{J_2}}\ton{t},~\upsigma_{\Delta\Omega^{\bds{\hat{S}}}_{J_2}}\ton{t}$ due to the uncertainties in the main constituents of the precession/nutation and of the change of the obliquity. Figure~\ref{figNUT} shows $\upsigma_{\Delta I^{\bds{\hat{S}}}_{J_2}}\ton{t},~\upsigma_{\Delta\Omega^{\bds{\hat{S}}}_{J_2}}\ton{t}$ over a time span 12 yr, from 2020 to 2032, for different values of the satellite's semimajor axis $a$. They were computed by using the nominal value of $J_2$ retrieved from some Earth's gravity model. It turned out that the magnitude of the initial offsets $\Delta I_0,~\Delta\Omega_0$ do not have a particular impact on the plots. It can be noted that, while the inclination reaches at most $\simeq ~0.08~\textrm{mas}$ for low-altitude orbits,  the node is slightly more sensitive to the mismodeling in the precession/nutation and the obliquity since its shift can be as large as $\simeq ~0.2~\textrm{mas}$ for $a = 7000~\textrm{km}$.
}
\section{The tidal perturbations}\lb{tides}
\subsection{the solid tides}\lb{soltid}
The long-term perturbations due to the solid component of the $\ell=2,~m=1,~p=1,~q=0$ constituent of the tesseral tide $K_1$  on the satellite's inclination and node, referred to the equator, are
\begin{align}
\dot I\eqt^{K_1,\textrm{s}} &= -\sqrt{\rp{5}{24\uppi}}\rp{3g_\oplus R_\oplus ^3k_{2,1,K_1}^{\ton{0}}H_2^1\ton{K_1}\cos I\eqt}{2\nk a^5}\sin\ton{\Omega\eqt - \delta_{2,1,K_1}}\lb{IsolK1}, \\ \nonumber \\
\dot\Omega\eqt^{K_1,\textrm{s}} &= \sqrt{\rp{5}{24\uppi}}\rp{3g_\oplus R_\oplus ^3k_{2,1,K_1}^{\ton{0}}H_2^1\ton{K_1}\ton{1-2\cos^2 I\eqt}}{2\nk a^5\sin I\eqt}\cos\ton{\Omega\eqt - \delta_{2,1,K_1}}.\lb{OsolK1}
\end{align}
They can be calculated by applying the Lagrange planetary equations for the rates of change of the inclination and the node \citep{2003ASSL..293.....B} to Eq.~(18) of \citet{2001CeMDA..79..201I}.
\textcolor{black}{In \rfr{IsolK1} and \rfr{OsolK1}, $g_\oplus$ is the Earth's acceleration of gravity at the equator, while $k_{2,1,K_1}^{\ton{0}},~H_2^1\ton{K_1},~\delta_{2,1,K_1}$ are the dimensionless frequency-dependent Love number, the solid tidal height, and the phase lag of the response of the solid Earth with respect to the constituent $K_1$ of degree $\ell=2$ and order $m=1$, respectively.}
According to \rfr{dotiec} \textcolor{black}{and} \rfr{dotOec}, the solid $K_1$-induced rate of changes of the inclination and node, referred to the ecliptic, are
\begin{align}
\dot I^{K_1,\textrm{s}} \nonumber &= \sqrt{\rp{5}{24\uppi}}\rp{3g_\oplus R_\oplus ^3k_{2,1,K_1}^{\ton{0}}H_2^1\ton{K_1}}{2\nk a^5}\qua{-\cos\epsilon\cos I\eqt\ton{\sin\Omega\eqt\cos\delta_{2,1,K_1} - \right.\right.\\ \nonumber \\
\nonumber &-\left.\left. \cos\Omega\eqt\sin\delta_{2,1,K_1}} +\sin\epsilon\ton{\rp{1-2\cos^2 I\eqt }{\sin I\eqt}}\ton{\cos\Omega\eqt\cos\delta_{2,1,K_1} + \right.\right.\\ \nonumber \\
&+\left.\left. \sin\Omega\eqt\sin\delta_{2,1,K_1}}    }\lb{solydaI},\\ \nonumber\\
\dot \Omega^{K_1,\textrm{s}} \nonumber &= \sqrt{\rp{5}{24\uppi}}\rp{3g_\oplus R_\oplus ^3k_{2,1,K_1}^{\ton{0}}H_2^1\ton{K_1}}{2\nk a^5}\qua{\sin\epsilon\cos I\eqt\ton{\sin\Omega\eqt\cos\delta_{2,1,K_1} - \right.\right.\\ \nonumber \\
\nonumber &-\left.\left. \cos\Omega\eqt\sin\delta_{2,1,K_1}} +\cos\epsilon\ton{\rp{1-2\cos^2 I\eqt }{\sin I\eqt}}\ton{\cos\Omega\eqt\cos\delta_{2,1,K_1} + \right.\right.\\ \nonumber \\
&+\left.\left. \sin\Omega\eqt\sin\delta_{2,1,K_1}}    }\lb{solydaO}
\end{align}
in which \rfrs{cosieq}{sinOeq} are to be used to express \rfr{solydaI} \textcolor{black}{and} \rfr{solydaO} in terms of the ecliptical orbital elements.

An alternative approach to straightforwardly obtain \rfr{solydaI} \textcolor{black}{and} \rfr{solydaO} consists of expressing the perturbing tidal potential of Eq.~(18) of \citet{2001CeMDA..79..201I} for the solid component of $K_1$ with $\ell = 2,~m = 1,~p = 1,~q = 0$ in terms of the ecliptical orbital elements and, then, applying the Lagrange planetary equations, which are not restricted to any coordinate system, to them. In this way, it is also possible to straightforwardly infer that, in the case $I = \Omega = 90\deg$, the  long-term perturbations due to the zonal constituent $055.565$ with $\ell = 2,~m = 0,~p = 1,~q = 0$ vanish for both the inclination and the node.

In the ideal case $e = 0,~I = \Omega = 90\deg$,
\rfr{solydaI} \textcolor{black}{and} \rfr{solydaO} become
\begin{align}
\dot I^{K_1,\textrm{s}} \lb{Is90}& =\sqrt{\rp{5}{24\uppi}}\rp{3g_\oplus R_\oplus ^3k_{2,1,K_1}^{\ton{0}}H_2^1\ton{K_1}\sin\delta_{2,1,K_1}\sin\epsilon}{2\nk a^5}, \\ \nonumber \\
\dot \Omega^{K_1,\textrm{s}} \lb{IO90}& =\sqrt{\rp{5}{24\uppi}}\rp{3g_\oplus R_\oplus ^3k_{2,1,K_1}^{\ton{0}}H_2^1\ton{K_1}\sin\delta_{2,1,K_1}\cos\epsilon}{2\nk a^5}\textcolor{black}{.}
\end{align}
\subsection{The ocean tides}\lb{octid}
As far as the ocean component of the $\ell=2,~m=1,~p=1,~q=0$ $K_1$ tidal line are concerned, we have
\begin{align}
\dot I\eqt^{K_1,\textrm{oc}} \lb{bibi}& = \rp{6 G\rho_\textrm{w} R_\oplus ^4 \ton{1+k_2^{'}}C^{+}_{2,1,K_1}\cos I\eqt}{5\nk a^5\ton{1-e^2}^2}\cos\ton{\Omega\eqt - \varepsilon^{+}_{2,1,K_1}}, \\ \nonumber \\
\dot \Omega\eqt^{K_1,\textrm{oc}} \lb{bibo}& = \rp{6 G\rho_\textrm{w} R_\oplus ^4 \ton{1+k_2^{'}}C^{+}_{2,1,K_1}\ton{1-2\cos^2 I\eqt}}{5\nk a^5\ton{1-e^2}^2\sin I\eqt}\sin\ton{\Omega\eqt - \varepsilon^{+}_{2,1,K_1}}.
\end{align}
They can be obtained with the Lagrange planetary equations applied to Eq.~(46) of \citet{2001CeMDA..79..201I}.
\textcolor{black}{In \rfr{bibi} and \rfr{bibo}, $\rho_\textrm{w}$ is the volumetric ocean water density, $k_2^{'}$ is the dimensionless load Love number, while $C_{2,1,K_1}^{+},~\varepsilon^{+}_{2,1,K_1}$ are the ocean tidal height  and the phase shift due to hydrodynamics of the oceans for the tidal constituent $K_1$ of degree $\ell=2$ and order $m=1$, respectively.}
From \rfr{dotiec} \textcolor{black}{and} \rfr{dotOec}, applied to \rfr{bibi} \textcolor{black}{and} \rfr{bibo}, one gets
\begin{align}
\dot I^{K_1,\textrm{oc}} \nonumber & = \rp{6 G\rho_\textrm{w} R_\oplus ^4 \ton{1+k_2^{'}}C^{+}_{2,1,K_1}}{5\nk a^5\ton{1-e^2}^2}\qua{\cos\epsilon\cos I\eqt\ton{\cos\Omega\eqt\cos\varepsilon^{+}_{2,1,K_1} + \sin\Omega\eqt\sin\varepsilon^{+}_{2,1,K_1}} +\right.\\ \nonumber\\
&+\left. \sin\epsilon\ton{\rp{1-2\cos^2 I\eqt}{\sin I\eqt}}\ton{\sin\Omega\eqt\cos\varepsilon^{+}_{2,1,K_1} - \cos\Omega\eqt\sin\varepsilon^{+}_{2,1,K_1}}  }  \lb{oceanaI}, \\ \nonumber \\
\dot \Omega^{K_1,\textrm{oc}} \nonumber & = \rp{6 G\rho_\textrm{w} R_\oplus ^4 \ton{1+k_2^{'}}C^{+}_{2,1,K_1}}{5\nk a^5\ton{1-e^2}^2}\qua{-\sin\epsilon\cos I\eqt\ton{\cos\Omega\eqt\cos\varepsilon^{+}_{2,1,K_1} + \sin\Omega\eqt\sin\varepsilon^{+}_{2,1,K_1}} +\right.\\ \nonumber\\
&+\left. \cos\epsilon\ton{\rp{1-2\cos^2 I\eqt}{\sin I\eqt}}\ton{\sin\Omega\eqt\cos\varepsilon^{+}_{2,1,K_1} - \cos\Omega\eqt\sin\varepsilon^{+}_{2,1,K_1}}  } \lb{oceanaO}.
\end{align}

Also \rfr{oceanaI} \textcolor{black}{and} \rfr{oceanaO} can be alternatively obtained by expressing Eq.~(46) of \citet{2001CeMDA..79..201I} for the ocean component of $K_1$ with $\ell = 2,~m = 1,~p = 1,~q = 0$ in terms of the ecliptical orbital elements and, then, using the Lagrange planetary equations.

For $I = \Omega = 90\deg$, \rfr{oceanaI} \textcolor{black}{and} \rfr{oceanaO} reduce to
\begin{align}
\dot I^{K_1,\textrm{oc}} \lb{Io90}& = \rp{6 G\rho_\textrm{w} R_\oplus ^4 \ton{1+k_2^{'}}C^{+}_{2,1,K_1}\cos\varepsilon^{+}_{2,1,K_1}\sin\epsilon}{5\nk a^5}, \\ \nonumber \\
\dot \Omega^{K_1,\textrm{oc}} \lb{Oo90}& = \rp{6 G\rho_\textrm{w} R_\oplus ^4 \ton{1+k_2^{'}}C^{+}_{2,1,K_1}\cos\varepsilon^{+}_{2,1,K_1}\cos\epsilon}{5\nk a^5}.
\end{align}
\subsection{The impact of the mismodeling in the tidal parameters}
\begin{figure*}
\centerline{
\vbox{
\begin{tabular}{cc}
\epsfysize= 5.0 cm\epsfbox{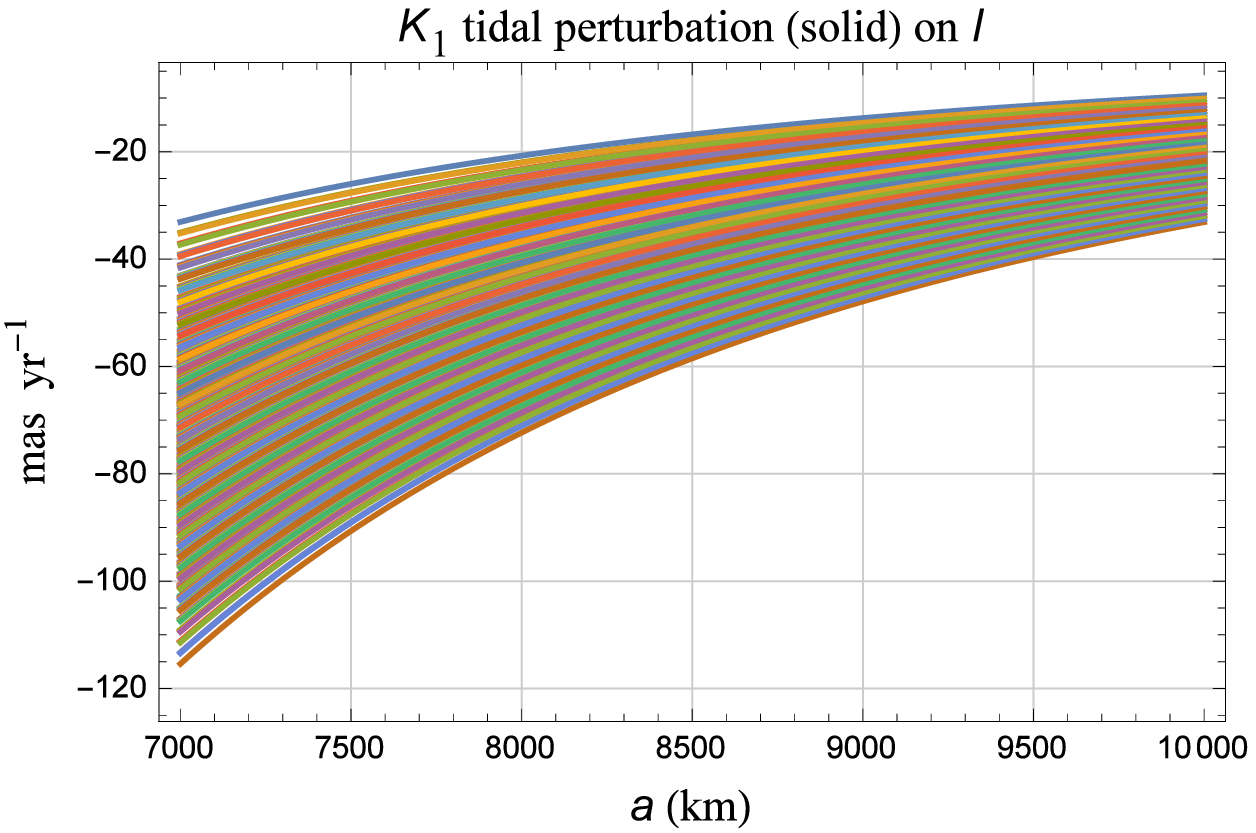} & \epsfysize= 5.0 cm\epsfbox{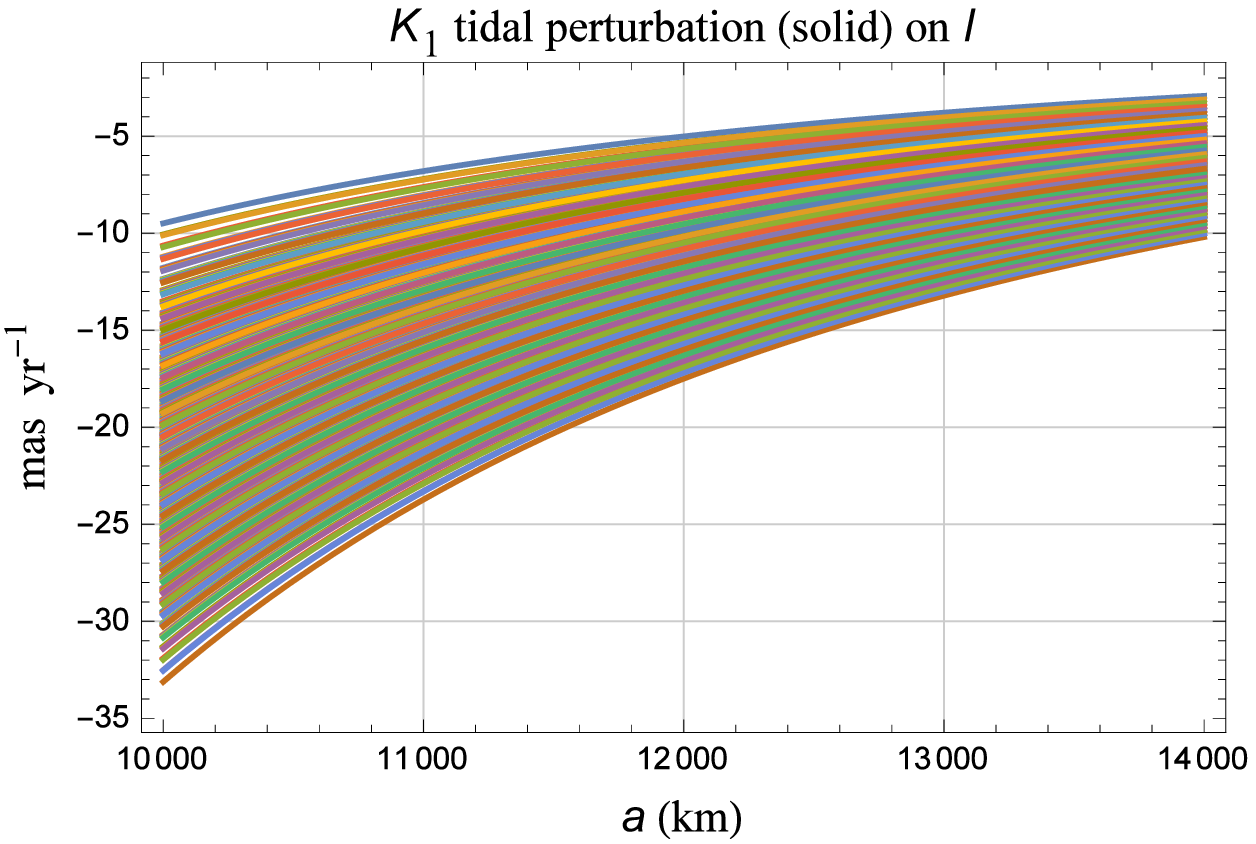}\\
\epsfysize= 5.0 cm\epsfbox{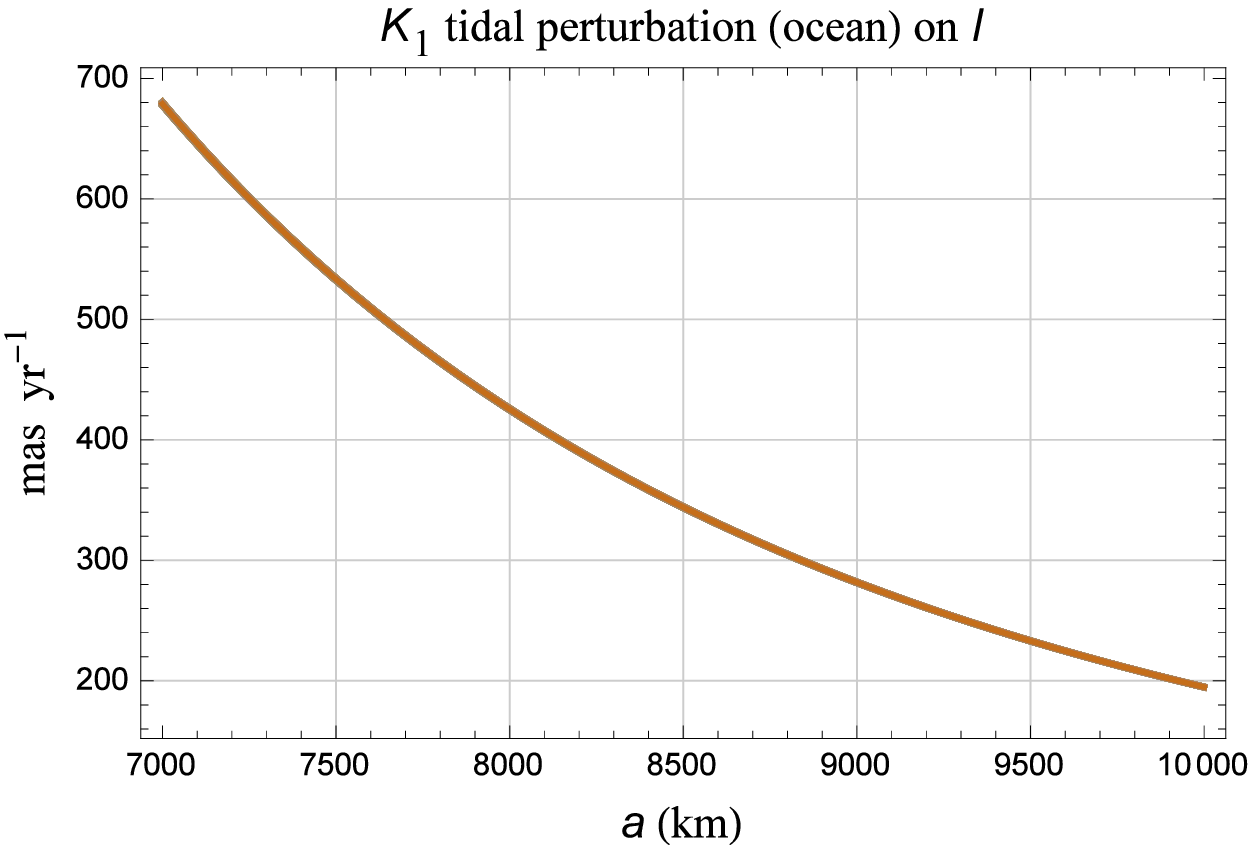} & \epsfysize= 5.0 cm\epsfbox{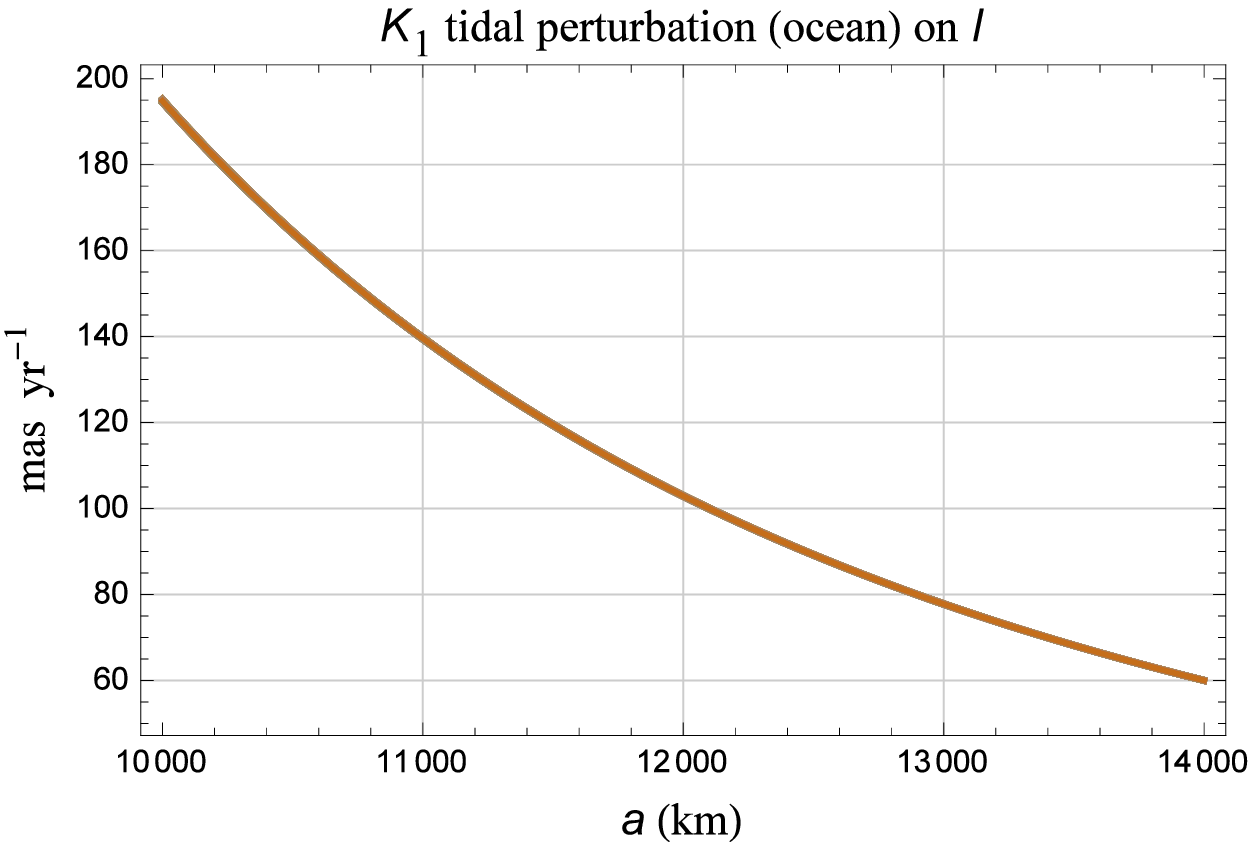}\\
\end{tabular}
}
}
\caption{Nominal perturbations  due to the solid (upper row) and ocean (lower row) component of the  $\ell=2,~m=1,~p=1,~q=0$ constituent of the $K_1$  tide on the satellite's inclination $I$ as functions of $a$, as per \rfr{solydaI} \textcolor{black}{and} \rfr{oceanaI}. In all the panels, each curve corresponds to a pair of values of $I,~\Omega$ within the ranges $I=\Omega=90\pm 0.05\deg$. The current level of uncertainty in the Love number $k_{2,1,K_1}$ is of the order of $\simeq 10^{-3}$ or, perhaps, one order of magnitude better \citep{polacchi018}. According to the past EMG96 model \citep{EGM96},  $C^{+}_{2,1,K_1}$ was known with a relative accuracy of $4\times 10^{-2}$. However, by calculating mean and standard deviation of  the values computed at  https://bowie.gsfc.nasa.gov/ggfc/tides/harmonics.html from the models TPXO.6.2 \citep{2002JAtOT..19..183E}, GOT99 \citep{got99} and  FES2004 \citep{2006OcDyn..56..394L}, a relative uncertainty of $1.8\times 10^{-3}$ is inferred. }\label{fig6}
\end{figure*}

\begin{figure*}
\centerline{
\vbox{
\begin{tabular}{cc}
\epsfysize= 5.0 cm\epsfbox{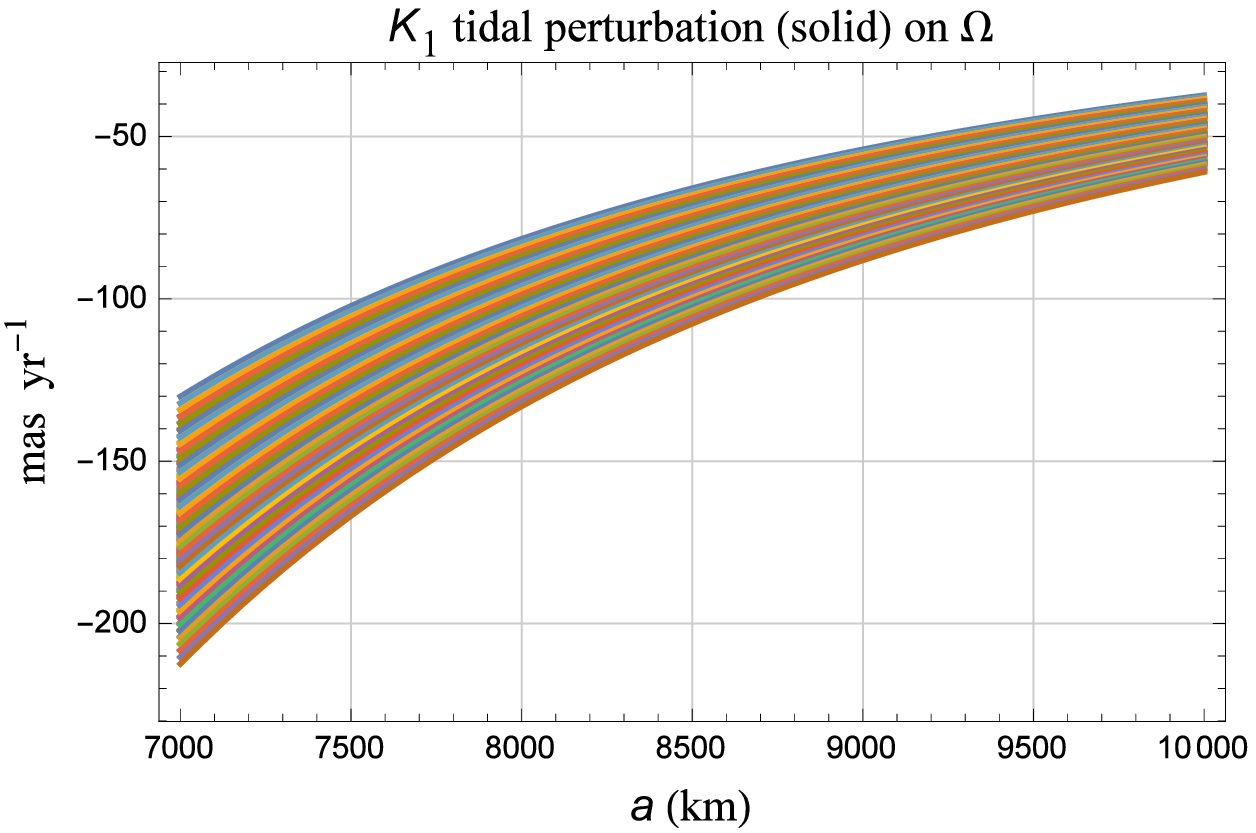} & \epsfysize= 5.0 cm\epsfbox{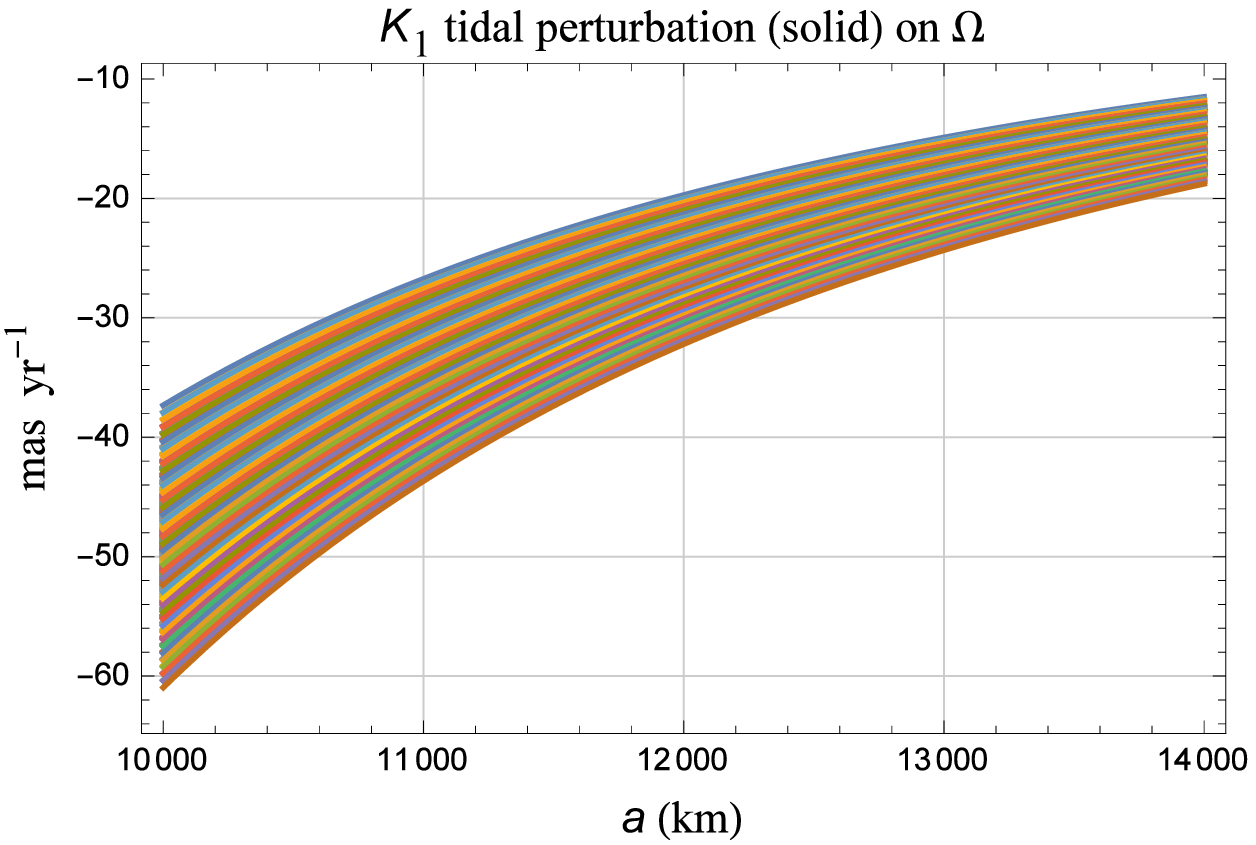}\\
\epsfysize= 5.0 cm\epsfbox{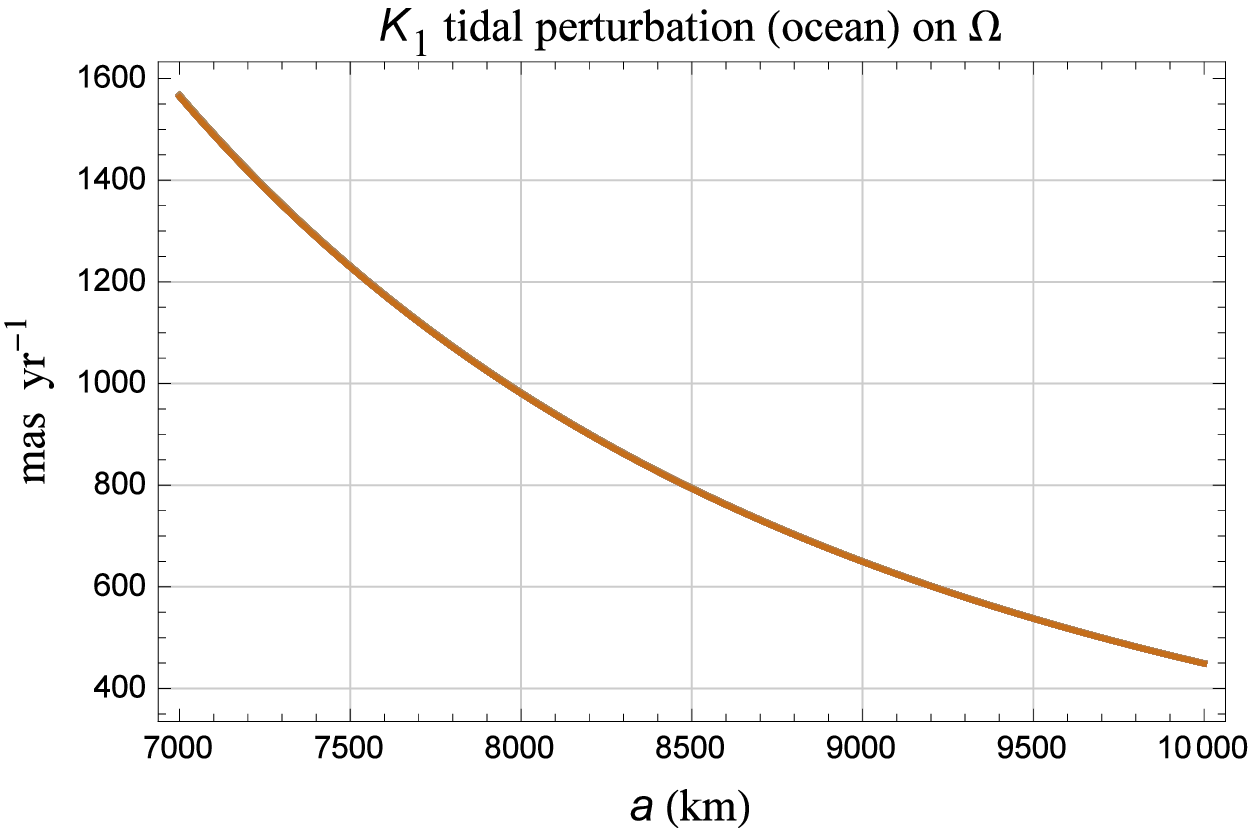} & \epsfysize= 5.0 cm\epsfbox{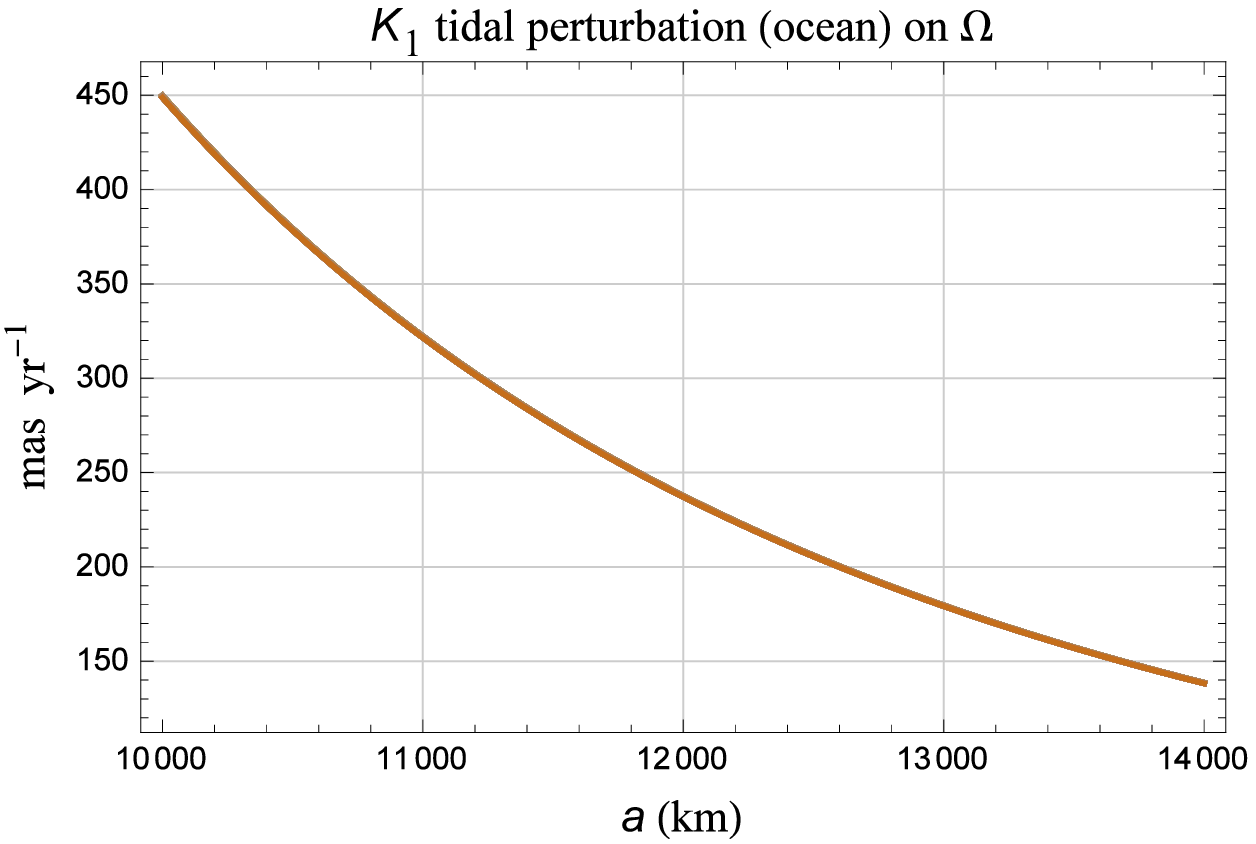}\\
\end{tabular}
}
}
\caption{Nominal perturbations  due to the solid (upper row) and ocean (lower row) component of the  $\ell=2,~m=1,~p=1,~q=0$ constituent of the $K_1$  tide on the satellite's node $\Omega$ as functions of $a$, as per \rfr{solydaO} \textcolor{black}{and} \rfr{oceanaO}. In all the panels, each curve corresponds to a pair of values of $I,~\Omega$ within the ranges $I=\Omega=90\pm 0.05\deg$. The current level of uncertainty in the Love number $k_{2,1,K_1}$ is of the order of $\simeq 10^{-3}$ or, perhaps, one order of magnitude better \citep{polacchi018}. According to the past EMG96 model \citep{EGM96},  $C^{+}_{2,1,K_1}$ was known with a relative accuracy of $4\times 10^{-2}$. However, by calculating mean and standard deviation of  the values computed at  https://bowie.gsfc.nasa.gov/ggfc/tides/harmonics.html from the models TPXO.6.2 \citep{2002JAtOT..19..183E}, GOT99 \citep{got99} and  FES2004 \citep{2006OcDyn..56..394L}, a relative uncertainty of $1.8\times 10^{-3}$ is inferred.}\label{fig7}
\end{figure*}

\lb{senstid}
Figure~\ref{fig6} \textcolor{black}{and} Figure~\ref{fig7} show the \textcolor{black}{sensitivity} of \rfr{solydaI} \textcolor{black}{and} \rfr{solydaO} and \rfr{oceanaI} \textcolor{black}{and} \rfr{oceanaO}, plotted as functions of the satellite's semimajor axis $a$, to departures of $I,~\Omega$ from the ideal configuration $I = \Omega = 90\deg$. By noting that the Love number $k_{2,1,K_1}$ seems currently known at an accuracy level not better than\footnote{L. Petrov and R. Ray, personal communications, August 2018. Nonetheless, in \citet{polacchi018} a relative uncertainty  as little as $3\times 10^{-4}$ was reported on a generic $k_2$ Love number determined with the LAGEOS and LAGEOS II satellites.} $\simeq 10^{-3}$, it turns out that not too large offsets $\Delta\Omega,~\Delta I\simeq 0.05\deg$ would be adequate to cope with the solid tidal perturbation; suffice it to say that, for GP-B, it was $\Delta I\eqt = 5\times 10^{-5}\deg$ at its launch \citep[p. 141]{gpbrep}. On the other hand, the nominal ocean tidal perturbations are larger than the solid ones;
such a discrepancy is due to the different values of their lag angles $\delta_{2,1,K_1},~\varepsilon^{+}_{2,1,K_1}$ so that, while $\sin \delta_{2,1,K_1} \simeq -0.005$, on the other hand it is $\cos\varepsilon^{+}_{2,1,K_1} \simeq 0.77$.
Thus, the present-day level of uncertainty in the ocean tidal height coefficient of $K_1$  is of crucial importance to assess the level of aliasing which could be induced on the relativistic signatures. If one had to rely upon the old EGM96 model along with its $4\times 10^{-2}$ relative uncertainty in  $C^{+}_{2,1,K_1}$ \citep{EGM96}, the bias on the Lense-Thirring signature would be at a $\simeq 40-50\%$ level.
However, several other global ocean tide models have been produced since then: CSR4.0 \citep{csr40} TPXO.6.2 \citep{2002JAtOT..19..183E} GOT99 \citep{got99}  FES2004 \citep{2006OcDyn..56..394L}, EOT11a \citep{EOT11a},  EOT11ag \citep{2012JGeo...59...28M}. By a comparison among them, it does not seem unrealistic to assume a present-day relative uncertainty of the order of $\simeq 10^{-3}$ for $C^{+}_{2,1,K_1}$. Indeed, by calculating mean and standard deviation of  the values computed at  https://bowie.gsfc.nasa.gov/ggfc/tides/harmonics.html from the models TPXO.6.2 \citep{2002JAtOT..19..183E}, GOT99 \citep{got99} and  FES2004 \citep{2006OcDyn..56..394L}, a relative uncertainty of $1.8\times 10^{-3}$ is inferred. It would yield an aliasing level of the Lense-Thirring signatures of a few percent.
\subsection{The linear combination approach}\lb{combtid}
At first sight, a possible way to overcome such an issue would consist of suitably designing a linear combination of the satellite's inclination and node which, by construction, cancels out both the ocean and solid tidal perturbations due to the $K_1$ line. By means of \rfr{Io90} \textcolor{black}{and} \rfr{Oo90} it is possible to obtain
\eqi
f \doteq \dot I + c_1\dot\Omega,\lb{combi}
\eqf
with
\eqi
c_1 = -\tan\epsilon = -0.433547.\lb{coef}
\eqf
Unfortunately, the linear combination of \rfr{combi} cancels out also the Lense-Thirring precessions; indeed,  \rfr{LT} \textcolor{black}{and} \rfr{LTO} reduce just to
\begin{align}
\dot I_\textrm{LT} & = \rp{2GS\sin\epsilon}{c^2 a^3}, \\ \nonumber \\
\dot \Omega_\textrm{LT} &= \rp{2GS\cos\epsilon}{c^2 a^3}
\end{align}
for $e = 0,~I = \Omega = 90\deg$.
It is a consequence of a general result about $f$ which can be drawn for $I = \Omega = 90\deg$  for the perturbations induced by any disturbing acceleration.
Indeed, from \rfr{dotiec} \textcolor{black}{and} \rfr{dotOec} and \rfr{combi} \textcolor{black}{and} \rfr{coef}, it turns out that the combined signature for the inclination and node rates of change due to a generic perturbing acceleration $A_\textrm{pert}$ of whatsoever physical origin, is
\eqi
f^\textrm{pert} = \rp{\dot I^\textrm{pert}\eqt}{\cos\epsilon},\lb{generale}
\eqf
where the analytical expression for $\dot I\eqt^\textrm{pert}$ has to be evaluated for $I = \Omega = 90\deg$.
In the case of the Lense-Thirring effect, \rfr{generale} tells us immediately that $f^\textrm{LT}$ vanishes since there is no gravitomagnetic precession for $I$ in the equatorial coordinate system.
\section{The De Sitter precessions}\lb{des}
On the other hand, \rfr{combi} \textcolor{black}{and} \rfr{coef} \textcolor{black}{have} the advantage of returning a non-vanishing effect due to the post-Newtonian gravitoelectric De Sitter precessions of the satellite's inclination and node.

\begin{figure*}
\centerline{
\vbox{
\begin{tabular}{cc}
\epsfysize= 5.0 cm\epsfbox{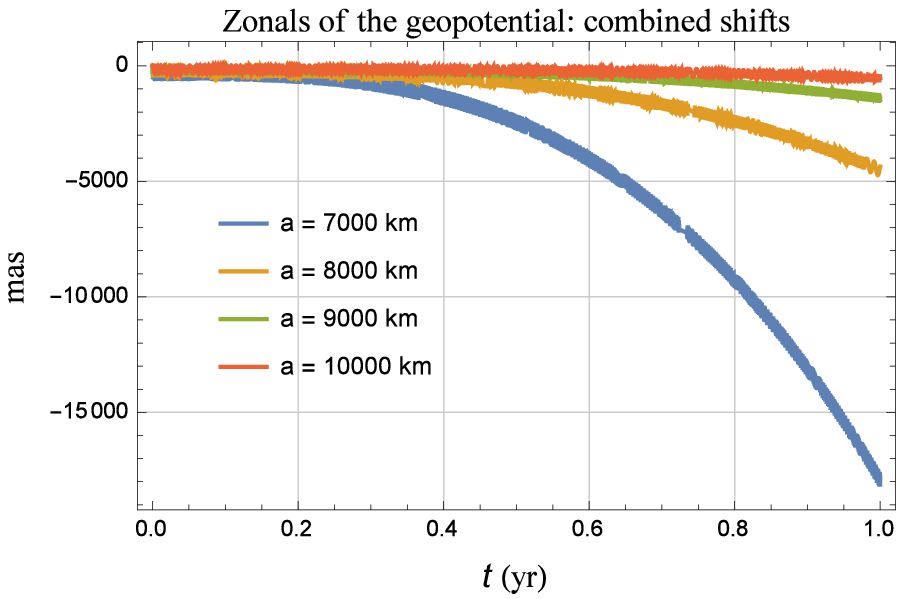} & \epsfysize= 5.0 cm\epsfbox{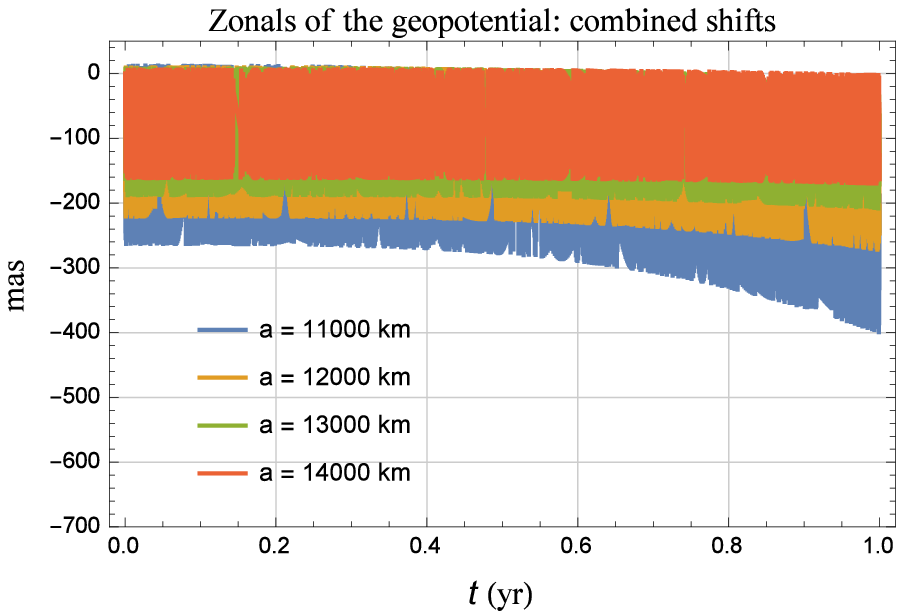}\\
\end{tabular}
}
}
\caption{Numerically generated nominal amplitudes of the precessions of the satellite's inclination and node induced by the first five  zonal harmonics $J_2,~J_3,~J_4,~J_5,~J_6$ of the geopotential linearly combined according to \rfr{combi} \textcolor{black}{and} \rfr{coef}. They were obtained, for different values of $a$, by subtracting two time series for the combination of \rfr{combi} \textcolor{black}{and} \rfr{coef} produced by numerically integrating the equations of motion in rectangular Cartesian coordinates with and without the classical accelerations due to $J_\ell,~\ell = 2,~3,~4,~5,~6$. Both the runs shared the same initial conditions characterized, among other things, by $e = 0,\Omega = I = 90\pm 0.1\deg$.  The largest contribution is due to $J_2$, whose present-day uncertainty may be as large as $\lesssim 2\times 10^{-10}$ if evaluated conservatively; the statistical, formal errors $\upsigma_{{\overline{C}}_{2,0}}$ released in the global gravity field models produced from the GRACE/GOCE data by several institutions around the world are even $\simeq 1-3$ orders of magnitude smaller. In both cases, a reference frame with the mean ecliptic at the epoch J2000.0 was used as reference $\grf{x,~y}$ plane so that $\kx=0,~\ky=\sin\epsilon = 0.3978,~\kz=\cos\epsilon = 0.9175$. }\label{fig8}
\end{figure*}

Indeed, by using Eq.~(4) and Eq.~(8) of \citet{2018arXiv180901730I}, which describe the geodetic rates of change of the satellite's inclination and node with respect to any\footnote{It is understood that it has to be intended as kinematically rotating \citep{1989NCimB.103...63B}.} coordinate system, in the scenario $I = \Omega =  90\deg$,
\rfr{combi} \textcolor{black}{and} \rfr{coef} return
\eqi
f^\textrm{DS} = -\rp{3 \mu_\odot \nk^\oplus \ton{\cos\Omega_\oplus \sin I_\oplus + \cos I_\oplus\tan\epsilon}}{2c^2 a_\oplus\ton{1-e^2_\oplus}} = -8.31986\masy.\lb{dScombi}
\eqf
\textcolor{black}{In \rfr{dScombi}, $\mu_\odot = GM_\odot$ is the Sun's gravitational parameter, $a_\oplus,~e_\oplus$ are the semimajor axis and the eccentricity of the heliocentrc Earth's orbit, respectively, $\nk^\oplus$ is the Keplerian mean motion of the Earth's orbit, while $I_\oplus,~\Omega_\oplus$ are the inclination and the node of the Earth's orbit referred to the ecliptic, respectively.}
The signature of \rfr{dScombi} is essentially a secular trend since the inclination and the node of the heliocentric Earth's orbit change over timescales of the order of $\simeq 0.1-1~\textrm{Myr}$, as can be inferred by their extremely slow rates of change \citep{2000ssd..book.....M}.
It is important to remark that \rfr{dScombi} would not be affected, by construction, by the largest and most insidious among the tidal perturbations, i.e. the $K_1$ tide; the zonal tide $055.565$ is of no concern since in Section~\ref{tides} it was shown that its long-term perturbations vanish for both the inclination and the node. Furthermore, also the static part of the geopotential would be of no concern, as previously shown in Section~\ref{prece} for the inclination and the nodes taken separately. Figure~\ref{fig8} shows the impact of departures $\Delta I = \Delta\Omega = 0.1\deg$ from the condition $I = \Omega =  90\deg$ on the nominal zonals perturbations combined according to \rfr{combi} \textcolor{black}{and} \rfr{coef}. The corresponding mismodeled signatures would be completely negligible even by assuming very conservative levels of uncertainty in $J_\ell,~\ell \geq 2$.
\textcolor{black}{
Figure~{comboNUT} shows that the uncertainties in the precession/nutation parameters and in the obliquity do not affect the combination of \rfr{combi}. Indeed, it turns out that, over 12 yr, the combined mismodeled shift ranges from just $0.003~\textrm{mas}$ ($a = 7000~\textrm{km}$) to $0.0006~\textrm{mas}$ ($a = 14000~\textrm{km}$).
}
\section{The 3rd-body perturbations: the Sun and the Moon}\lb{3rd}
The results of \citet{2012CeMDA.112..117I} concerning the rates of change of the satellite's orbital elements, averaged over its orbital period $\Pb$, induced by a distant perturber X can be straightforwardly used in the present context since they are valid in any coordinate system.
The doubly averaged rate of change of the node can be obtained by averaging Eq.~(9) of \citet{2012CeMDA.112..117I} over the orbital period $P_\textrm{X}$ of the 3rd-body. The general result is
\begin{align}
\dot\Omega^\textrm{3rd body}\nonumber & = -\rp{3\GMX}{8 \aX^3 \sqrt{1 - e^2} (1 - \eX^2)^{3/2} \nk}\qua{ \cos\IX \cot I + \cos\ton{\Omega - \OX} \sin\IX}\times\\ \nonumber \\
\nonumber &\times \grf{-\ton{-2 - 3 e^2 + 5 e^2\cos 2\omega} \qua{\cos\IX \sin I -\cos I \cos\ton{\Omega - \OX} \sin\IX} -\right.\\ \nonumber\\
&-\left. 5 e^2 \sin\IX \sin 2\omega \sin\ton{\Omega - \OX} }\lb{OX}.
\end{align}
\textcolor{black}{In \rfr{OX}, $\omega$ is the argument of pericenter of the satellite's orbit referred to the ecliptic, while $\GMX,~\aX,~\eX,~\IX,~\OX$ are the gravitational parameter, the semimajor axis, the eccentricity, the inclination, the node of the body X, respectively. In the present context, $\IX,~\OX$ are to be meant as referred to the ecliptic.}
Eq.~(20) of \citet{2018arXiv180901730I} and \rfr{OX}, for $e = 0,~I = \Omega = 90\deg$, reduce to
\begin{align}
\dot I^\textrm{3rd body}\lb{uno}&= \rp{3\GMX\sin^2\IX \sin 2\OX}{8\aX^3\ton{1-\eX^2}^{3/2}\nk}, \\ \nonumber \\
\dot\Omega^\textrm{3rd body}\lb{due}&= -\rp{3\GMX \sin 2\IX\sin\OX}{8\aX^3\ton{1-\eX^2}^{3/2}\nk},
\end{align}
which can be linearly combined according to \rfr{combi} \textcolor{black}{and} \rfr{coef} giving
\eqi
f^\textrm{3rd body} = \rp{3\GMX \sin\IX\sin\OX\ton{\sin\IX\cos\OX +\tan\epsilon\cos\IX }}{4\aX^3\ton{1-\eX^2}^{3/2}\nk}.\lb{combi3rd}
\eqf

As far as the Moon is concerned, its node \textcolor{black}{$\Omega_{\leftmoon}$}, referred to the ecliptic, undergoes a secular precession with a period of $T_{\Omega_{\leftmoon}} = 18.6~\textrm{yr}$; its current value is $\Omega_{\leftmoon} = 125.1\deg$. Thus, although at the price of a long wait, \rfrs{uno}{combi3rd} will finally average out in view of their  frequencies $\dot\Omega_{\leftmoon},~2\dot\Omega_{\leftmoon}$, and it can be stated that they would be of no concern for either the individual Lense-Thirring precessions of \rfr{LT} \textcolor{black}{and} \rfr{LTO} and the combined De Sitter effect of \rfr{dScombi}.
Such a conclusion is true for each of the gravitomagnetic signatures also without waiting for the completion of a full cycle of the lunar node, as it can be easily checked by inspecting the maximum values of the mismodeled parts of \rfr{uno} \textcolor{black}{and} \rfr{due} for any value of $a$ and comparing them to Figure~\ref{fig1} \textcolor{black}{and} Figure~\ref{fig2}. To this aim, we assumed a relative uncertainty in the selenocentric gravitational parameter \textcolor{black}{$\mu_{\leftmoon}=GM_{\leftmoon}$} of $2\times 10^{-8}$, as per the Object Data Page of the Moon provided by the JPL HORIZONS Web interface, revised on 2013.
The situation is subtler for the combined De Sitter trend in view of the higher level of accuracy pursued. The maximum impact on \rfr{combi3rd} occurs when $\Omega_{\leftmoon} = 90\deg$, so that
\eqi
f_\textrm{max}^{\leftmoon} = \rp{3GM_{\leftmoon} \tan\epsilon\sin 2 I_{\leftmoon}}{8 a_{\leftmoon}^3\ton{1 - e_{\leftmoon}^2}^{3/2}\nk}.\lb{maxi3rd}
\eqf
\textcolor{black}{In \rfr{maxi3rd}, $I_{\leftmoon}$ is the inclination of the geocentric lunar orbit to the ecliptic, while $a_{\leftmoon},~e_{\leftmoon}$ are the semimajor axis and the eccentricity of the geocentric Moon's orbit, respectively.}
In Figure~\ref{fig9} we plot the mismodeled part of \rfr{maxi3rd} as a function of $a$. It can be noted that
it stays in the range $\simeq 3-10\times 10^{-5}$ of \rfr{dScombi}.

About the 3-body effect of the Sun, it is completely negligible. Indeed, it is
\eqi
f_\textrm{max}^\odot \simeq \rp{M_\odot}{M_{\leftmoon}}\ton{\rp{a_{\leftmoon}}{a_\oplus}}^3\rp{\sin 2 I_\oplus}{\sin 2 I_{\leftmoon}} f_\textrm{max}^{\leftmoon}= 3\times 10^{-4}~f_\textrm{max}^{\leftmoon}.
\eqf
Furthermore, the heliocentric gravitational parameter is currently known with a relative accuracy of   $7\times 10^{-11}$ \citep{2015JPCRD..44c1210P}.
\section{The non-gravitational perturbations}\lb{ngp}
The long-term rates of change induced by the non-gravitational accelerations on the inclination and node, referred to the equator, of the LAGEOS-type satellites  have been extensively investigated in the literature in the context of their bias on the equator-referred Lense-Thirring node precession by finding them at the percent level or less for LAGEOS. Thus, it is arguable that the same holds also when they are combined according to \rfr{dotiec} \textcolor{black}{and} \rfr{dotOec}  in order to yield their ecliptical counterparts. Indeed, the averaged node perturbations which vanish for $e=0,~I=\Omega=90\deg$ are those due to the Poynting-Robertson effect \citep[p.~ 608]{2016MNRAS.460..802L}, the infrared radiation pressure of the Earth \citep[p.~176]{1981CeMec..25..169S}, the atmospheric drag \citep[p.~103]{Nobilibook87}, the hypothetical asymmetric reflectivity \citep[Eq.~(44),~p.~1083]{2002P&SS...50.1067L}, the geomagnetic field, as can be inferred by integrating the Gauss equation for the rate of change of the node by means of \citet[Eq.~(24),~p.~592]{2014RAA....14..589A}, with $1/\sin f$ in its first term corrected to $\sin f$, to zero order in $e$ for $I=90\deg$, the secular trend due to the Yarkovsky-Rubincam effect \citep[Eq.~(19),~p.~1075]{2002P&SS...50.1067L}, the Earth albedo \citep[Eq.~(35),~p.~456]{2001P&SS...49..447L} and the solar radiation pressure \citep[Eq.~(18),~p.~451]{2001P&SS...49..447L} in absence of eclipses. In the case of eclipses, by using  the first term in the series of Eq.~(2) and Eq.~(4) in \citet{1972CeMec...5...80F} for the shadow function it can be shown that, to zero order in $e$, the node rates due to the albedo and the solar radiation pressure do not vanish; nonetheless, they turn out to be proportional to $\sin 2\uplambda_\odot$, which averages out after $1/2~\textrm{yr}$. About the long-term signatures due to the Yarkovsky-Rubincam effect having the frequencies $\dot\Omega$ \citep[Eq.~(18),~p.~1075]{2002P&SS...50.1067L}  and $2\dot\Omega$ \citep[Eq.~(17),~p.~1075]{2002P&SS...50.1067L}, they vanish if the satellite's spin axis is perpendicular to the Earth's equator. The same holds for the secular rate due to the Yarkovsky-Schach effect \citep[Eq.~(34),~p.~1079]{2002P&SS...50.1067L}. A similar situation occurs for the inclination, as shown in Sec.~(6) of  \citet{2018arXiv180901730I}.

As far as the De Sitter effect is concerned, by recalling the general result of \rfr{generale}, all the findings of Sec.~(6) of  \citet{2018arXiv180901730I} retain their validity, pointing towards an impact of the non-gravitational perturbations on the combined De Sitter trend of \rfr{dScombi}
globally meeting our requirement.

\textcolor{black}{We stress that the aforementioned considerations are to be deemed just as very preliminary because they are based on previous results with the existing LAGEOS type satellites. To this aim, it is important to remark that the architecture of, say, LAGEOS and LAGEOS II is a very old one and have several drawbacks. Just to mention a few of them, in view of the particular distribution of their corner cube retroreflectors (CCR), in practice, the satellites do not behave like a perfect sphere. Moreover, the virtual reflection point is not the center of mass of each satellite. A forthcoming, dedicated paper will be devoted to their consideration for the proposed new spacecraft ELXIS. This is particularly important for those thermal thrust forces which need the knowledge of the spin rate and orientation of the probe. }
\section{A comparison with the counter-orbiting scenario by van Patten and Everitt}\lb{vanpa}
Some decades ago, it was proposed by van Patten and Everitt (vPE) to launch a pair of counter-orbiting drag-free spacecraft in nearly identical circular and polar orbits at an altitude of about $800~\textrm{km}$ to perform a $\simeq 1\%$ measurement of both the Lense-Thirring and De Sitter node precessions by monitoring the sum of their nodes  \citep{1976PhRvL..36..629V,1976CeMec..13..429V,1977JSpRo..14..474S,1978AcAau...5...77V}.
In addition to the drag-free apparatus to counteract the non-gravitational perturbations, the vPE's satellites should have been endowed also with the capability of reciprocal Doppler tracking at mutual encounters when passing over the poles. Careful arrangements to avoid in-orbit collisions would have been required as well \citep{1976JAnSc..24..137S}.
The main, striking differences between ELXIS and the vPE's proposal are as follows
\begin{description}
  \item[$\left. a\right)$] The ELXIS concept is based on just a single satellite instead of two spacecraft, as in the vPE's proposal.
 \begin{description}
 \item[$\left. a^{I}\right)$] ELXIS should not be necessarily too complex and/or expensive since a comparatively simpler, well manufactured cannonball geodetic satellite of LAGEOS-type would likely fit our accuracy requirements since, as we have demonstrated, most of the non-gravitational perturbations vanish or average out after more or less long temporal intervals. Moreover, once in orbit, ELXIS could well wait for forthcoming improvements in both the tracking accuracy and in dynamical modeling of, e.g., the ocean tides, thus by allowing for repeated tests with likely continuously improved accuracy over the subsequent years.
 \item[$\left. a^{II}\right)$] There are no particular limitations on the orbital height, which can be conveniently set according to the unavoidable engineering/budgetary trade-off.
 \item[$\left. a^{III}\right)$] There are no collision-related issues.
 \item[$\left. a^{IV}\right)$] No careful satellite-to-satellite measurements of the angle between the two orbital planes at the poles are present.
 \item[$\left. a^{V}\right)$] The overall data analysis process of ELXIS would be much easier and less expensive than that encompassing two satellites\textcolor{black}{.}
\end{description}
\item[$\left. b\right)$] While \citet{1976PhRvL..36..629V} proposed to make a combined Lense-Thirring + De Sitter test, ELXIS would allow to perform separate-and even redundant-measurements of such two general relativistic effects. The De Sitter effect can be disentangled from the Lense-Thirring one both in the equatorial and in the ecliptical coordinate systems. Indeed, in the first case \citep{2018arXiv180901730I}, based on the analysis of the satellite's inclination only, the Lense-Thirring rate of change of it vanishes for $\kx = \ky = 0$, while in the second case, the linear combination
    of \rfr{combi} \textcolor{black}{and} \rfr{coef} cancels out just the gravitomagnetic precessions and enforces the geodetic ones yielding \rfr{dScombi}.
    Conversely, the Lense-Thirring measurements could be easily made independently of the geodetic effect itself simply by using an ecliptical version of the standard kinematically non-rotating and dynamically rotating geocentric coordinate system in which the De Sitter precession is automatically accounted for, not showing up in spacecraft motions. Basically, it would be just a geometrically rotated version of the usual ICRS adopted for routinely analyzing Earth's satellites data.
\item[$\left. c\right)$] With ELXIS, the accuracy in measuring the De Sitter combined precessions, independently of the Lense-Thirring effect, would be of the order of $\simeq 10^{-5}$, while \citep{1976PhRvL..36..629V} claimed a $\simeq 1\%$ accuracy in a mixed measurement of both the Lense-Thirring and the geodetic precessions. On the other hand, \citet{1976CeMec..13..429V} wrote that they could have obtained an independent test of the De Sitter effect at $10\%$, i.e. about 4 orders of magnitude worse than what could be obtained with ELXIS.
\item[$\left. d\right)$] The orbit injection errors on the nodes $\Omega$ of the vPE's satellites should have been of the order of $0.03\deg$ \citep{1976CeMec..13..429V}, which is a figure comparable with the ELXIS case, although offsets up to $0.1\deg$ would not compromise our accuracy goals. On the other hand, the requirements by \citet{1976CeMec..13..429V} on the inclinations of the orbital planes would have been of the order of $\simeq 0.0008\deg$, while for ELXIS they are at the $\simeq 0.01-0.1\deg$ level.
\item[$\left. e\right)$] A major drawback of the error budget of   \citet{1976PhRvL..36..629V,1976CeMec..13..429V} is, perhaps, that they seemingly did not take into account the perturbations due to the ocean tides induced on the satellites' motions by the free space tidally distorted Earth's potential.
    It is difficult to think that, say, the $K_1$ tide may have had no substantial effects on the vPE's spacecraft in view of the results obtained in the present study and of the fact that, at the time of \citet{1976PhRvL..36..629V,1976CeMec..13..429V}, the first, relatively rudimentary satellite-based  global ocean tide models, if any, are much less accurate than now. Suffice it to say that \citet{1976JGR....81.2557F} released errors in ocean tidal height $C_{2,1,K_1}^{+}$ as large as $\simeq 8-28\%$. Moreover, the uncertainty in it from the global Earth's gravity field model GEM-T3S \citep{1992gmef.book.....L}, published about 15 yr after the vPE's proposal, was still $6\%$. As a consequence, the claimed $\simeq 1-10\%$ accuracy levels claimed by \citet{1976PhRvL..36..629V,1976CeMec..13..429V} might have been somewhat optimistic.
\end{description}
Thus, it is clear that, even by limiting ourselves to the point $\left. a\right)$ and assuming hypothetically the same accuracy goals, ELXIS should be deemed as more advantageous than the vPE's proposal. Moreover, even by pessimistically surmising that our evaluation of the actually obtainable  accuracy on the De Sitter measurement should be moved to, say, the $\simeq 10^{-4}$ level, it would still be 3 orders of magnitude better than $10\%$ \citep{1976CeMec..13..429V}.
\section{Summary and conclusions}\lb{fine}
Using a geocentric kinematically rotating ecliptical coordinate system to analyze the data of a single Earth's satellite, provisionally named ELXIS, placed in a circular orbit perpendicular to the equator and to the reference direction of the Vernal Equinox has the advantage of allowing to use both its inclination and node to measure the general relativistic Lense-Thirring and De Sitter effects without being impacted by the competing classical long-term precessions due to the even and odd zonals of the geopotential, which ideally vanish. On the other hand, the ocean component of the $K_1$ tide induces aliasing perturbations which may degrade the accuracy of the proposed recovery of the individual gravitomagnetic and geodetic signatures. Depending on the actual mismodeling in the global ocean tide models, their impact on each of the Lense-Thirring rates may be at the percent level, representing the most limiting factor in measuring them. It is not unrealistic to expect further improvements in the forthcoming global tide solutions; in any case, once the satellite is in orbit, it would always be possible to wait just enough for the tidal models to reach the required accuracy. Moreover, it should be recalled that, after all, the accuracy of GP-B is $19\%$\textcolor{black}{. Other tests of the gravitomagnetic field of the Earth performed with the existing LAGEOS type satellites have been reported in the literature, but, contrary to GP-B, they are still somewhat controversial, especially as far as the actual accuracy reached}. As far as the De Sitter effect is concerned, \textcolor{black}{currently known at the $\simeq 10^{-3}$ level from LLR and GP-B,} the goal of a $\simeq 10^{-5}$ accuracy of its measurement can be enforced with respect to an equatorial frame by linearly combining its geodetic precessions on the satellite's inclination and node in such a way to cancel out, by construction, the solid and tidal perturbations due to the $K_1$ and other tides. The impact of the combined 3rd-body perturbations due to the Moon, which are relevant only for the De Sitter test at the considered level of accuracy, averages out after $18.6~\textrm{yr}$ since the smallest characteristic frequency of their combined signatures is just that of the lunar node. However, given the current level of uncertainty in the selenocentric gravitational parameter, the largest bias during a full cycle of it would be no larger than $\simeq 3\times 10^{-5}-1\times 10^{-4}$. The non-gravitational perturbations, preliminarily examined by assuming a cannonball geodetic satellite of LAGEOS-type, should not be a concern. \textcolor{black}{However, these are just  preliminary guesses based on past studies in the literature performed for the existing LAGEOS type satellites, whose physical structure is quite old and pose certain drawbacks which should be overcome with an entirely new spacecraft. A dedicated, forthcoming paper will be devoted to the non-conservative forces acting on ELXIS, especially the thermal thrusts which crucially depend on the spin rate and orientation of the spacecraft. } Such conclusions hold substantially for any value of the satellite's semimajor axis and for departures as large as $\simeq 0.01-0.1\deg$ from the ideal orbital geometry proposed.
\textcolor{black}{Finally, we note that  we worked in an analytical way by using the Keplerian orbital elements for the sake of simplicity and clarity about the rationale of the proposed experiment, and in order to offer to the reader a preliminary error budget easy to understand. If, on the one hand, long time series of some orbital elements have been actually used so far in the performed tests of the Lense-Thirring effect with the LAGEOS type satellites, on the other hand, a more robust and reliable approach would consist of explicitly modeling the features of motion one is interested in and estimating  one or more dedicated solve-for parameters in the data reduction by inspecting their correlations with the other determined parameters in the full covariance matrix. It is hoped that the present paper will boost further investigations by other researchers who may want to complement it with full covariance analyses implying numerical data simulations and reductions for, say, one or more types of spacecraft.}
\section*{Acknowledgements}
I am grateful to S. Kopeikin for some important clarifications and to an anonymous referee for the idea of comparing the present proposal with a past one.
\textcolor{black}{Also the critical remarks by three other anonymous referees were much appreciated}.
\begin{appendices}
\section{Notations and definitions}\lb{appena}
Here, some basic notations and definitions used in the text are presented \textcolor{black}{\citep{1991ercm.book.....B,2003ASSL..293.....B,2003A&A...412..567C,2010ITN....36....1P,2011rcms.book.....K,2014grav.book.....P,2017A&A...597A..83L}}. For the numerical values of some of them, see Tables~\ref{tavola0}\textcolor{black}{~to~\ref{nutt}}. The orbital elements referred to the mean Earth's equator at the reference epoch J2000.0 are denoted with the subscript \virg{eq} in the main text.
\begin{description}
\item[] $G:$ Newtonian constant of gravitation
\item[] $c:$ speed of light in vacuum
\item[] $\epsilon:$ mean obliquity
\item[] $\epsilon_0:$ mean obliquity at the reference epoch J2000.0
\item[] $\dot\epsilon:$ secular rate of the mean obliquity
\item[] $\zeta,~\vartheta,~z:$ precession angles
\item[] $\zeta_0,~z_0:$ precession angles at the reference epoch J2000.0
\item[] $\dot\zeta,~\dot\vartheta,~\dot z:$ secular rates of the precession angles
\item[] $\Delta\psi,~\Delta\epsilon:$ nutation angles
\item[] $A_{\Delta\psi_{\Omega_{\leftmoon}}},~A_{\Delta\epsilon_{\Omega_{\leftmoon}}}:$ amplitudes of the largest components of the nutation angles $\Delta\psi,~\Delta\epsilon$ due to the $18.6$-yr lunar motion
\item[] $M_\oplus:$ mass of the Earth
\item[] $\mu_\oplus\doteq GM_\odot:$ gravitational parameter of the Earth
\item[] $S_\oplus:$ magnitude of the angular momentum of Earth
\item[] ${\bds{\hat{S}}}_\oplus = \grf{0,~\sin\epsilon,~\cos\epsilon}:$ spin axis of the Earth in an ecliptic coordinate system
\item[] $R_\oplus:$ equatorial radius of the Earth
\item[] ${\overline{C}}_{\ell,m}:$ fully normalized Stokes coefficient of degree $\ell$ and order $m$ of the  multipolar expansion of the Earth's gravitational potential
\item[] $J_\ell=-\sqrt{2\ell+1}~{\overline{C}}_{\ell,0}:$ zonal harmonic coefficient of degree $\ell$ of the multipolar expansion of the Earth's gravitational potential
\item[] $g_\oplus:$ Earth's acceleration of gravity at the equator
\item[] $k_{2,1,K_1}^{\ton{0}}:$ dimensionless frequency-dependent Love number for the $K_1$ tidal constituent of degree $\ell=2$ and order $m=1$
\item[] $H_2^1\ton{K_1}:$ frequency-dependent solid tidal height for the $K_1$ constituent of degree $\ell=2$ and order $m=1$
\item[] $\delta_{2,1,K_1}:$ phase lag of the response of the solid Earth with respect to the constituent $K_1$ of degree $\ell=2$ and order $m=1$.
\item[] $\rho_\textrm{w}:$ volumetric ocean water density
\item[] $k_2^{'}:$ dimensionless load Love number
\item[] $C_{2,1,K_1}^{+}:$ ocean tidal height  for the constituent $K_1$ of degree $\ell=2$ and order $m=1$.
\item[] $\varepsilon^{+}_{2,1,K_1}:$ phase shift due to hydrodynamics of the oceans for the tidal constituent $K_1$ of degree $\ell=2$ and order $m=1$.
\item[] $\bds r:$ satellite's position vector with respect to the Earth
\item[] $r:$ magnitude of the satellite's position vector with respect to the Earth
\item[] $a:$  semimajor axis of the geocentric satellite's orbit
\item[] $\nk \doteq \sqrt{\mu_\oplus a^{-3}}:$  Keplerian mean motion of the geocentric satellite's orbit
\item[] $\Pb\doteq 2\uppi\nk^{-1}:$ orbital period of the geocentric satellite's orbit
\item[] $e:$  eccentricity of the geocentric satellite's orbit
\item[] $I:$  inclination of the orbital plane of the geocentric satellite's orbit to the mean ecliptic at the reference epoch J2000.0
\item[] $\Omega:$  longitude of the ascending node  of the geocentric satellite's orbit referred to the mean ecliptic at the reference epoch J2000.0
\item[] \textcolor{black}{$\omega:$  argument of perigee of the geocentric satellite's orbit referred to the mean ecliptic at the reference epoch J2000.0}
\item[] $\bds{\hat{n}} = \grf{\sin I\sin\Omega,~-\sin I\cos\Omega,~\cos I}:$ normal unit vector in an ecliptic coordinate system. It is perpendicular to the satellite's orbital plane
\item[] $M_\odot:$ mass of the Sun
\item[] $\mu_\odot\doteq GM_\odot:$ gravitational parameter of the Sun
\item[] $\lambda_\odot:$ ecliptic longitude of the Sun
\item[] $a_\oplus:$  semimajor axis of the heliocentric Earth's orbit
\item[] $n_\textrm{b}^\oplus \doteq \sqrt{\mu_\odot a_\oplus^{-3}}:$  Keplerian mean motion of the heliocentric Earth's orbit
\item[] $P_\oplus\doteq 2\uppi{n_\textrm{b}^\oplus}^{-1}:$ orbital period of the heliocentric Earth's orbit
\item[] $e_\oplus:$  eccentricity of the heliocentric Earth's orbit
\item[] $I_\oplus:$  inclination of the orbital plane of the heliocentric Earth's orbit to the mean ecliptic at the reference epoch J2000.0
\item[] $\Omega_\oplus:$  longitude of the ascending node  of the heliocentric Earth's orbit referred to the mean ecliptic at the reference epoch J2000.0
\item[] $M_\textrm{X}:$ mass of the 3rd body X (Sun $\odot$ or Moon $\leftmoon$)
\item[] $\mu_\textrm{X}\doteq GM_\textrm{X}:$ gravitational parameter of the 3rd body X (Sun $\odot$ or Moon $\leftmoon$)
\item[] $a_\textrm{X}:$  semimajor axis of the geocentric orbit of the 3rd body X
\item[] $P_\textrm{X}:$ orbital period of the geocentric orbit of the 3rd body X
\item[] $e_\textrm{X}:$  eccentricity of the geocentric orbit of the 3rd body X
\item[] $I_\textrm{X}:$  inclination of the orbital plane of the geocentric orbit of the 3rd body X to the mean ecliptic at the reference epoch J2000.0
\item[] $\Omega_\textrm{X}:$  longitude of the ascending node  of the geocentric orbit of the 3rd body X referred to the mean ecliptic at the reference epoch J2000.0
\textcolor{black}{
\item[] $\dot\Omega_\textrm{X}:$  secular rate of the longitude of the ascending node  of the geocentric orbit of the 3rd body X referred to the mean ecliptic at the reference epoch J2000.0
}
\item[] $T_{\Omega_{\leftmoon}}:$ period of the node of the geocentric Moon's orbit referred to the mean ecliptic at the reference epoch J2000.0
\end{description}
\section{Tables and Figures}\lb{appenb}
\begin{table*}
\caption{Relevant physical and orbital parameters used in the text. Most of the reported values come from \citet{2001CeMDA..79..201I,2010ITN....36....1P} and references therein. The source for the orbital elements characterizing the heliocentric orbit of the Earth, referred to the mean ecliptic at the reference epoch J2000.0,  is the freely consultable database JPL HORIZONS on the Internet at https://ssd.jpl.nasa.gov/?horizons from which they were retrieved by choosing the time of writing this paper as input epoch.
For the level of accuracy with which some of the parameters listed here are currently known, see the main text.
}\lb{tavola0}
\begin{center}
\begin{tabular}{|l|l|l|}
  \hline
Parameter  & Units & Numerical value \\
\hline
$G$ & $\textrm{kg}^{-1}~\textrm{m}^3~\textrm{s}^{-2}$ & $6.67259\times 10^{-11} $\\
$c$ & $\textrm{m~s}^{-1}$ & $2.99792458\times 10^8$\\
%
%
%
$\mu_\oplus$ & $\textrm{m}^3~\textrm{s}^{-2}$ & $ 3.986004418\times 10^{14}$ \\
$S_\oplus$ & $\textrm{kg~m}^2~\textrm{s}^{-1}$ & $5.86\times 10^{33}$ \\
$R_\oplus$ & $\textrm{m}$ & $6.3781366\times 10^{6}$\\
%
%
${\overline{C}}_{2,0}$ & $-$ & $-4.84165299806\times 10^{-4}$\\
$g_\oplus$ & $\textrm{m~s}^{-2}$ & $9.7803278$\\
$k^{(0)}_{2,1,{K_1}}$ & $-$ & $0.257$\\
$H^1_2\ton{K_1}$ & $\textrm{m}$ & $0.3687012$\\
$\delta_{2,1,{K_1}}$ & $\textrm{deg}$ & $-0.3$\\
$\rho_\textrm{w}$ & $\textrm{kg~m}^{-3}$ & $1.025\times 10^3$ \\
$k^{'}_2$ & $-$ & $-0.3075$\\
$C^{+}_{2,1,{K_1}}$ & $\textrm{m}$ & $2.23659\times 10^{-2}$\\
$\varepsilon^{+}_{2,1,{K_1}}$ & $\textrm{deg}$ & $315.958$\\
%
%
$\mu_\odot$ & $\textrm{m}^3~\textrm{s}^{-2}$ & $ 1.32712440018\times 10^{20}$ \\
$a_\oplus$ & $\textrm{au}$ & $1.000003360971446$\\
$e_\oplus$ & $-$ & $0.01636541170625853$ \\
$I_\oplus$ & $\textrm{deg}$ & $0.003786566401597615$ \\
$\dot I_\oplus$ & $\textrm{deg~cty}^{-1}$ & $-0.01337178$\\
$\Omega_\oplus$ & $\textrm{deg}$ & $171.6446280787646$ \\
$\dot\Omega_\oplus$ & $\textrm{deg~cty}^{-1}$ & $-0.24123856$ \\
%
%
$\mu_{\leftmoon}$ & $\mu_\oplus$ & $1.23000371\times 10^{-2}$ \\
$a_{\leftmoon}$ & $\textrm{km}$ & $385,734$ \\
$e_{\leftmoon}$ & $-$ & $0.05183692147447081$\\
$I_{\leftmoon}$ & $\textrm{deg}$ & $5.208682439763778$ \\
$\Omega_{\leftmoon}$ & $\textrm{deg}$ & $125.1041727302047$\\
$\dot\Omega_{\leftmoon}$ & $\mathrm{\prime\prime}~\textrm{cty}^{-1}$ $-6,962,890.5431$ \\
$T_{\Omega_{\leftmoon}}$ & $\textrm{yr}$ & $\simeq 18.6$\\
%
\hline
\end{tabular}
\end{center}
\end{table*}
\begin{table*}
\caption{
\textcolor{black}{Numerical values and associated uncertainties of the relevant astronomical parameters of the precession/nutation and of the obliquity. They were retrieved from \citet{2017A&A...597A..83L} ($\epsilon_0,~\dot\epsilon$), \citet{2003A&A...412..567C} ($\zeta_0,~\dot\zeta,~\dot\vartheta,~z_0,~\dot z$), \citet{2010ITN....36....1P} (the amplitudes $A_{\Delta\psi_{\Omega_{\leftmoon}}},~A_{\Delta\epsilon_{\Omega_{\leftmoon}}}$ of the lunar node terms in $\Delta\psi,\Delta\epsilon$).}
}\lb{nutt}
\begin{center}
\textcolor{black}{
\begin{tabular}{|l|l|l|l|}
  \hline
Parameter  & Units & Numerical value & Uncertainty\\
\hline
$\epsilon_0$ & $\mathrm{\prime\prime}$ & $84,381.411365$ & $8\times 10^{-6}$ \\
$\dot\epsilon$ &  $\mathrm{\prime\prime}~\textrm{cty}^{-1}$ & $-460.0836735$ & $1\times 10^{-7}$ \\
$\zeta_0$ & $\mathrm{\prime\prime}$ & $2.650545$ & $1\times 10^{-6}$ \\
$\dot\zeta$ &  $\mathrm{\prime\prime}~\textrm{cty}^{-1}$ & $2,306.083227$ & $1\times 10^{-6}$ \\
$\dot\vartheta$ &  $\mathrm{\prime\prime}~\textrm{cty}^{-1}$ & $2,004.191903$ & $1\times 10^{-6}$ \\
$z_0$ & $\mathrm{\prime\prime}$ & $-2.650545$ & $1\times 10^{-6}$ \\
$\dot z$ &  $\mathrm{\prime\prime}~\textrm{cty}^{-1}$ & $2,306.077181$ & $1\times 10^{-6}$ \\
$A_{\Delta\psi_{\Omega_{\leftmoon}}}$ & $\upmu\textrm{as}$ & $-17,206,424.18$ & $1\times 10^{-2}$\\
$A_{\Delta\epsilon_{\Omega_{\leftmoon}}}$ & $\upmu\textrm{as}$ & $9,205,233.10$ & $1\times 10^{-2}$\\
\hline
\end{tabular}
}
\end{center}
\end{table*}
\begin{figure*}
\centerline{
\vbox{
\begin{tabular}{cc}
\epsfysize= 5.0 cm\epsfbox{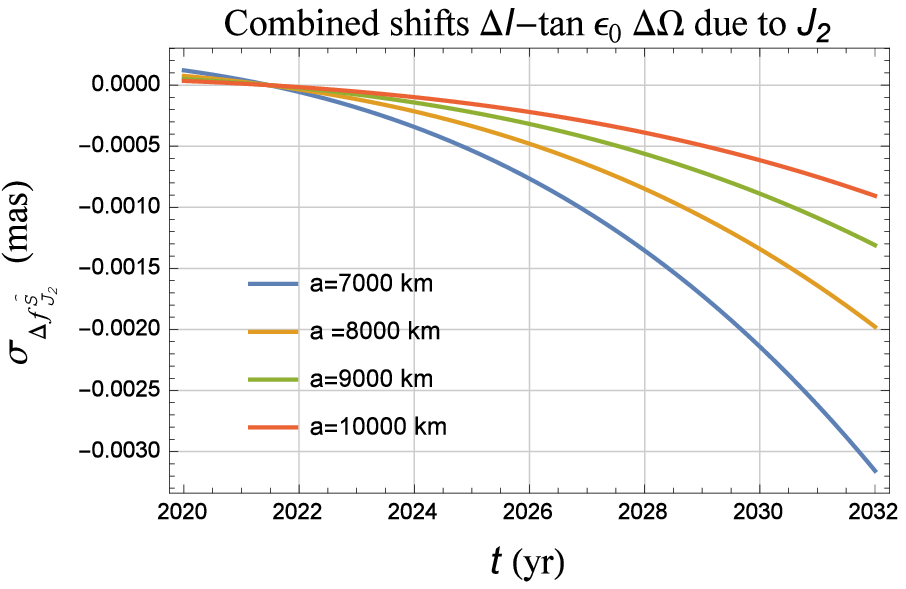} & \epsfysize= 5.0 cm\epsfbox{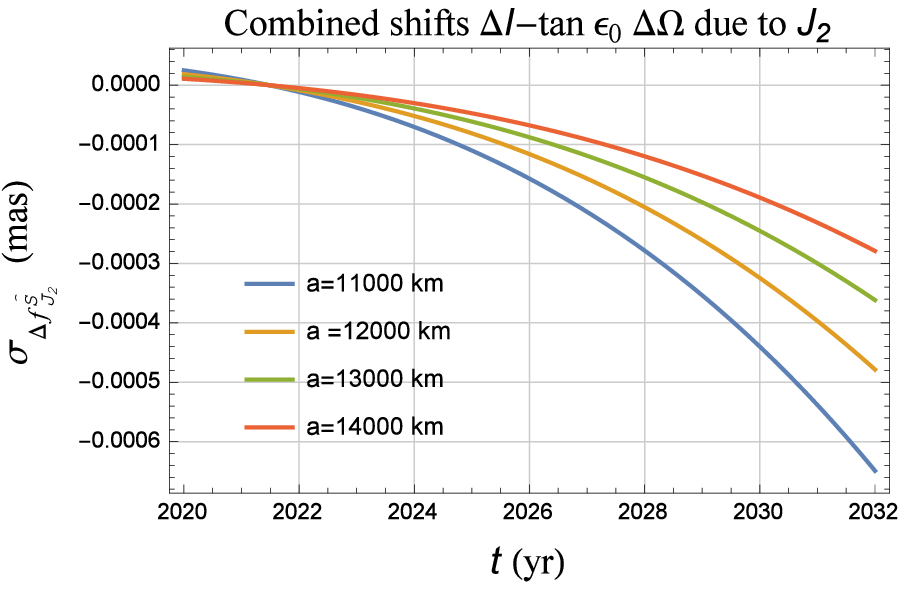}\\
\end{tabular}
}
}
\caption{\textcolor{black}{
Mismodeled combined time series $\upsigma_{\Delta f_{J_2}^{\bds{\hat{S}}}}\ton{t}$, produced analytically from \citet[Eqs.~(12)~to~(15)]{2011PhRvD..84l4001I} and \rfr{combi} and \rfr{coef}  for different values of the semimajor axis $a$ and $I=\Omega = 90 \pm 0.01\deg$, of the time-dependent $J_2$-induced shifts of the inclination $I$ and the node $\Omega$ due to the uncertainties of the parameters entering the precession/nutation and the temporal change of the obliquity according to the values listed in
Table~\ref{nutt} of Appendix~\ref{appenb}. The nominal value of $J_2$, retrieved from some model, was adopted.
}}\label{comboNUT}
\end{figure*}
\begin{figure*}
\centerline{
\vbox{
\begin{tabular}{cc}
\epsfysize= 5.0 cm\epsfbox{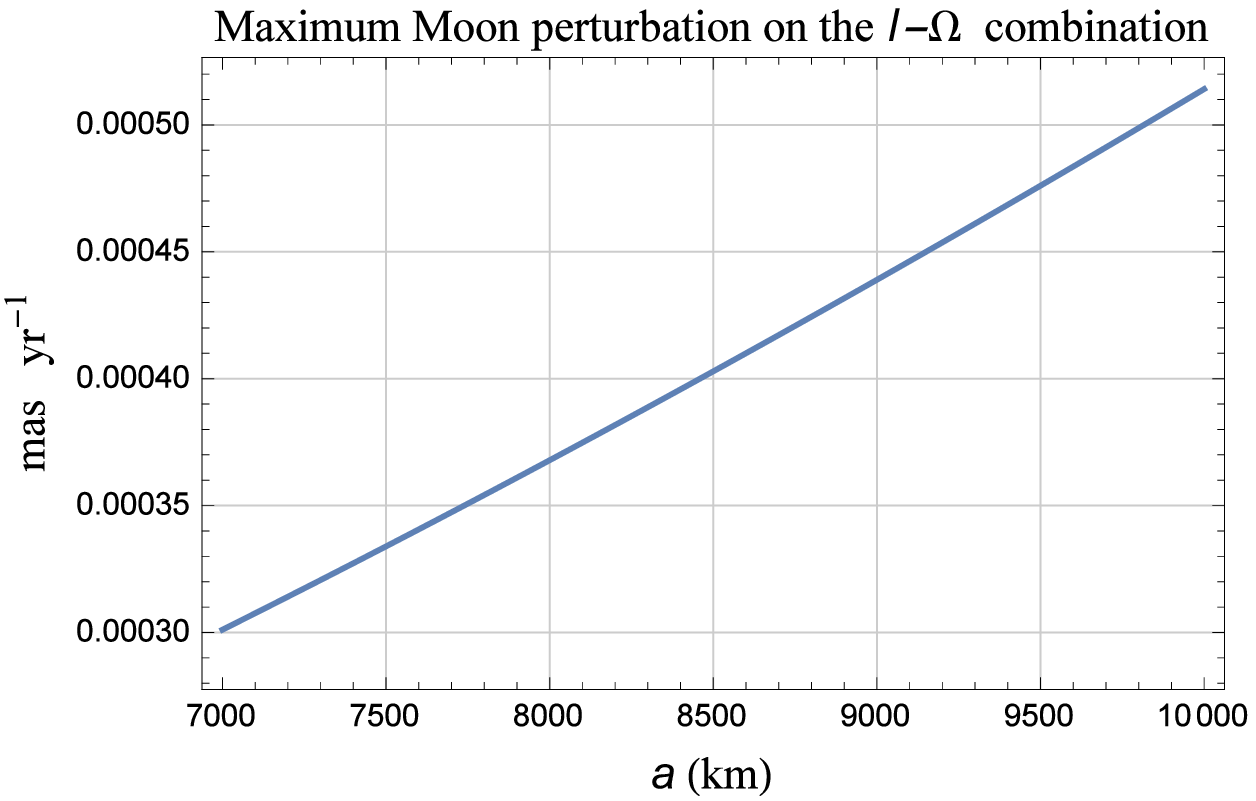} & \epsfysize= 5.0 cm\epsfbox{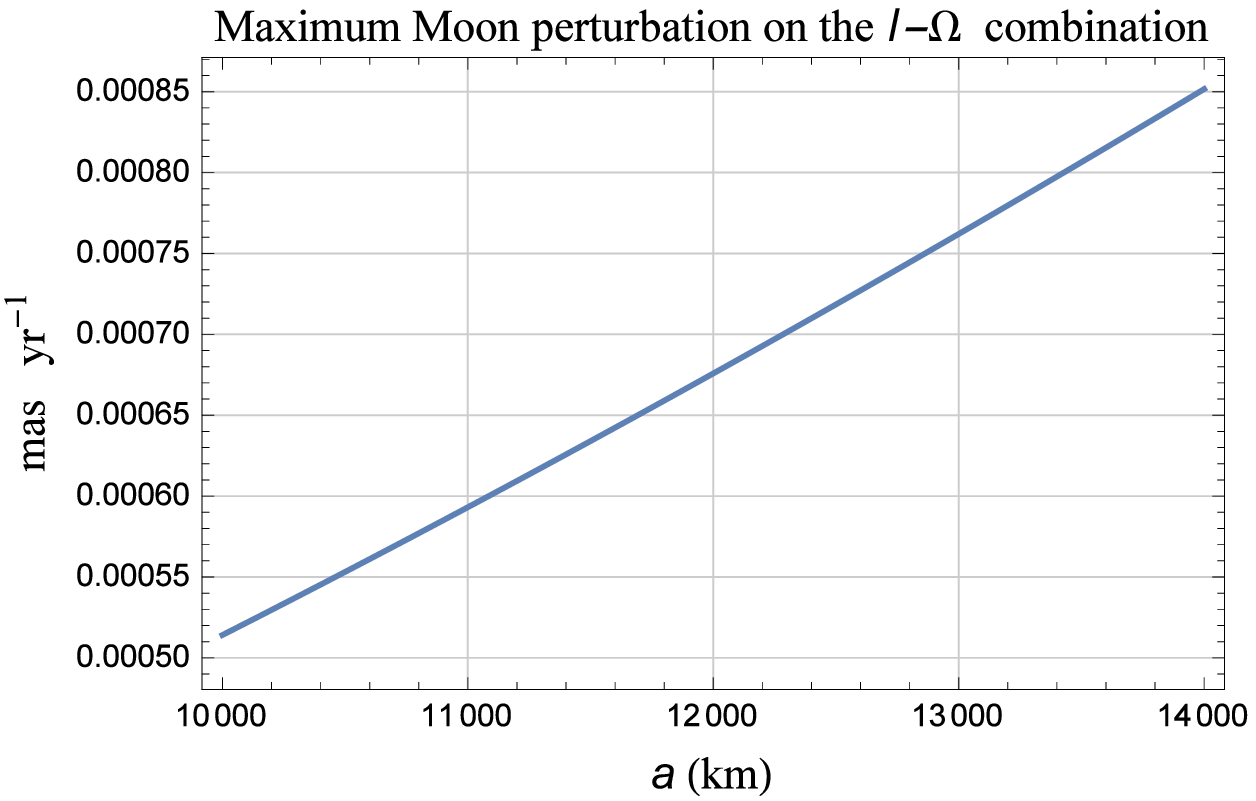}\\
\end{tabular}
}
}
\caption{Maximum value, in $\masy$, of the mismodelled part of the 3rd-body precessions of $I,~\Omega$ due to the Moon combined with \rfr{combi} \textcolor{black}{and} \rfr{coef} as a function of the satellite's semimajor axis $a$; \rfr{maxi3rd} was used. We assumed a relative uncertainty in $\mu_{\leftmoon}$ of $2\times 10^{-8}$, as per the Object Data Page of the Moon provided by the JPL HORIZONS Web interface, revised on 2013.}\label{fig9}
\end{figure*}
\end{appendices}
\bibliography{Gclockbib,semimabib,PXbib}{}


\end{document}